# High-throughput Design of Magnetic Materials

Hongbin Zhang
*Institute of Materials Science, TU Darmstadt, 64287 Darmstadt, Germany*

Materials design based on density functional theory (DFT) calculations is an emergent field of great potential to accelerate the development and employment of novel materials. Magnetic materials play an essential role in green energy applications as they provide efficient ways of harvesting, converting, and utilizing energy. In this review, after a brief introduction to the major functionalities of magnetic materials, we demonstrated the fundamental properties which can be tackled via high-throughput DFT calculations, with a particular focus on the current challenges and feasible solutions. Successful case studies are summarized on several classes of magnetic materials, followed by bird-view perspectives for the future.

**Contents**

**Glossary of acronyms**

2D: two-dimensional

AFM: antiferromagnetic

AHC: anomalous Hall conductivity

ANC: anomalous Nernst conductivity

AMR: anisotropic magnetoresistance

ARPES: angle-resolved photoemission

BZ: Brillouin zone

DFT: density functional theory

DLM: disordered local moment

DOS: density of states

DMFT: dynamical mean field theory

DMI: Dzyaloshinskii-Moriya interaction

DMS: dilute magnetic semiconductors

FiM: ferrimagnetic

FL: Fermi liquid

FM: ferromagnetic

FMR: ferromagnetic resonance

FOPT: first-order phase transition

GMR: giant magnetoresistance

HM: half-metal

HMFM: half-metallic ferromagnet

HTP: high-throughput

ICSD: inorganic crystal structure database

iSGE: inverse spin Galvanic effect

IFM: itinerant ferromagnet

IM: itinerant magnet

MAE: magnetocrystalline anisotropy energy

MBE: molecular beam epitaxy

MCE: magnetocaloric effect

MGI: materials genome initiative

ML: machine learning

$M_r$: remanent magnetization

$M_s$: saturation magnetization

MOKE: magneto-optical Kerr effect

MRAM: magnetic random-access memory

MSMA: magnetic shape memory alloy

MSMA: magnetic shape memory effect

NFL: non-Fermi liquid

NM: nonmagnetic

QAHE: quantum anomalous Hall effect

QAHI: quantum anomalous Hall insulator

QCP: quantum critical point

QPI: quasipartical interference

QPT: quantum phase transition

QSHE: quantum spin Hall effect

RE: rare-earth

RPA: random phase approximation

SCR: self-consistent renormalization

SdH: Shubnikov-de Haas

SDW: spin density wave

SGS: spin gapless semiconductor

SHE: spin Hall effect

SIC: self-interaction correction

SOC: spin-orbit coupling

SOT: spin-orbit torque

STS: scanning tunnelling spectroscopy

SQS: special quasi-random structure

STT: spin-transfer torque

$T_C$: Curie temperature

TI: topological insulator

TRIM: time-reversal invariant momenta

TM: transition metal

TMR: tunnelling magnetoresistance

$T_N$: Néel temperature

vdW: van der Waals

XMCD: x-ray magnetic circular dichroism

## I. INTRODUCTION

Advanced materials play an essential role in the functioning and welfare of the society, particularly magnetic materials as one class of functional materials susceptible to external magnetic, electrical, and mechanical stimuli. Such materials have a vast spectrum of applications thus are indispensable to resolve the current energy issue. For instance, permanent magnets can be applied for energy harvesting (*e.g.*, wind turbine to generate electricity) and energy conversion (*e.g.*, electric vehicles and robotics with mechanical energies from electricity). According to the BCC research report,[1] the global market for soft and permanent magnets reached $32.2 billion in 2016 and will reach $51.7 billion by 2022. Furthermore, to go beyond the quantum limit of conventional electronic devices, spintronics exploiting the spin degree of freedom of electrons provides a promising alternative for energy efficient apparatus, which has attracted intensive attention in the last decades. Nonetheless, there are still a variety of pending fundamental problems and emergent phenomena to be understood. Therefore, there is a strong impetus to develop better understanding of magnetism and magnetic materials, and to design magnetic materials with optimal performance.

The conventional way of discovering and employing materials is mostly based on the empirical structure-property relationships and try-and-error experiments, which are time and resource costly. Early in 2011, the U.S. government has launched the Materials Genome Initiative (MGI), aiming at strategically exploring materials design.[2] The proposed synergistic paradigm integrating theory, modelling, and experiment has proven to be successful, while there are still plenty of open challenges.[3] From the theoretical point of view, as the material properties comprise the intrinsic (as given by the crystal structure) and extrinsic (as given by the microstructure) contributions, a multi-scale modelling framework should be established and embraced, leading to the integrated computational materials engineering approach[4] and the European materials modelling council program.[5] For both frameworks and the counterparts, density functional theory (DFT) is of vital importance, due to its capability to obtain accurate electronic structure and thus the intrinsic properties, and essential parameters for multi-scale modelling, ensuring the predictive power.

Till now, high-throughput (HTP) computations based on DFT have been applied to screening for various functional materials, such as electro-catalyts,[6] thermoelectrics,[7] and so on. Correspondingly, open databases such as Materials Project,[8] AFLOWlib,[9] NOMAD,[10] and OQMD[11] have been established, with integrated platforms like AiiDA[12] and Atomate[13] available. This changes the way of performing DFT calculations from monitoring jobs on a few compounds to defining and applying workflows applicable on thousands of compounds, so that the desired properties get evaluated and optimized, *e.g.*, the thermoelectric figure of merit.[7]

In contrast to the other physical properties such as band gaps, absorption energies for catalysts, and thermoelectric properties, magnetic properties and their characterization based on DFT pose a series of unique challenges and there has been limited exploration of designing functional magnetic materials. For instance, there are three key intrinsic magnetic properties, *i.e.*, magnetization, magnetic anisotropy energy (MAE), and the critical ordering temperature, which are difficult to be evaluated in a HTP manner. Besides the intriguing origin of magnetization (*e.g.*, localized or itinerant) where consistent treatment requires a universal theoretical framework beyond local and semilocal approximations to DFT, it is already a tricky problem to identify the magnetic ground states as the magnetic moments can get ordered in ferromagnetic (FM), ferrimagnetic (FiM), antiferromagnetic (AFM), and even incommensurate noncollinear configurations. Moreover, the dominant contribution to the MAE can be attributed to the relativistic effects, *e.g.*, spin-orbit coupling (SOC), where the accurate evaluation demands good convergence with respect to the $k$-mesh, resulting in expensive computational efforts. Lastly, the critical ordering temperature is driven by the magnetic excitations and the corresponding thermodynamic properties cannot in principle be addressed by the standard DFT without extension to finite temperature. Due to such challenges, to the best of our knowledge, there are only a few limited successful stories about applying the HTP method on designing magnetic materials (cf. Sect. V for details).

In this review, we aim at illustrating the pending problems and discussing possible solutions to facilitate HTP design of magnetic materials, focusing particularly on the intrinsic properties for crystalline compounds which can be evaluated based on DFT calculations. Also, we prefer to draw mind maps with a priority on the conceptual aspects rather than the technical and numerical details, which will be referred to the relevant literature. Correspondingly, the major applications of magnets and the criticality aspects will be briefly summarized in Sect. II. In Sect. III, we will elucidate several fundamental aspects which are essential for proper HTP computations on magnetic materials, with detailed discussions on a few representative classes of magnetic materials. The in-depth mathematical/physical justification will not be repeated but referred to necessary publications for curious readers, *e.g.*, the recently published "Handbook of Magnetism and Advanced Magnetic Materials"[14] is a good source for elaborated discussions on the physics of magnetic materials. In Sect. V the successful case studies will be summarized, with future perspectives given in Sect. VI.

It is noted that there are a big variety of magnetic materials which are with intriguing physics or promising for applications, as listed in the magnetism roadmap.[15–17] We excuse ourselves for leaving out the current discussions on quantum magnets,[18–20] quantum spin liquid,[21] (curvilinear) nanomagnets,[22,23] Skyrmions,[24] high entropy alloys,[25] multiferroics,[26] and ultrafast magnetism,[27] which will be deferred to reviews specified. Also, we apologize for possible ignorance of specific publications due to the constraint on the man power.

## II. MAIN APPLICATIONS OF MAGNETIC MATERIALS

### A. Main applications

The main applications of magnetic materials are compiled in Table.I, together with the representative materials. We note that small magnetic nanoparticles (2-100 nm in diameter) can also be used in biology and medicine for imaging, diagnostics, and therapy,[28] which is beyond the scope of this review. Most remaining applications are energy related. For instance, both permanent magnets (*i.e.*, hard magnets) and soft magnets are used to convert the mechanical energy to electrical energy and *vice versa.* Whereas spintronics stands for novel devices operating with the spin of electrons, which are in principle more energy efficient than the established electronic devices based on semiconductors.

Permanent magnets are magnetic materials with significant magnetization (either FM or FiM) which are mostly applied in generating magnetic flux in a gap, with the corresponding figure of merit being the maximum energy product $(BH)_{max}$.[29] As marked by the B-H curves in Fig. 1, it provides an estimation of the energy stored in the magnets and can be enhanced by maximizing the hysteresis, *e.g.*, by increasing the remanent magnetization $M_r$ and coercivity $H_c$. Such extrinsic quantities like $M_r$ and $H_c$ are closely related to the microstructure of the materials, where the corresponding upper limits are given by the intrinsic quantities such as saturation magnetization $M_s$ and MAE. Nowadays, for commercially available $Nd_2Fe_{14}B$, 90% of the theoretical limit of $(BH)_{max}$ can be achieved,[30] suggesting that there is a substantial space to further improve the performance of other permanent magnets.

In contrast, soft magnets are magnetic materials with significant $M_s$ as well, and they are easy to be magnetized and demagnetized corresponding to high initial and maximal relative permeability $\mu_r = B/(\mu_0 H)$, where $B = \mu_0(H+M)$ denotes the magnetic induction under magnetic field H, and $\mu_0$ is the vacuum magnetic permeability. That is, the corresponding hysteresis loop as shown in Fig. 1 is ideally narrow for soft magnets, corresponding to vanishing hysteresis in the ideal case. In this regard, most soft magnets are Fe-based alloys of cubic structures, because Fe has the largest average moment among the $3d$ series of elements in solids. The soft magnets are widely applied in power generation, transmission, and distribution (Table. I). In addition to the $M_s$ which is an intrinsic property, microstructures play a significant role in developing soft magnets, *e.g.*, the eddy current losses caused by the cyclical rearrangements of magnetic domains in AC magnetic fields should be minimized.[31]

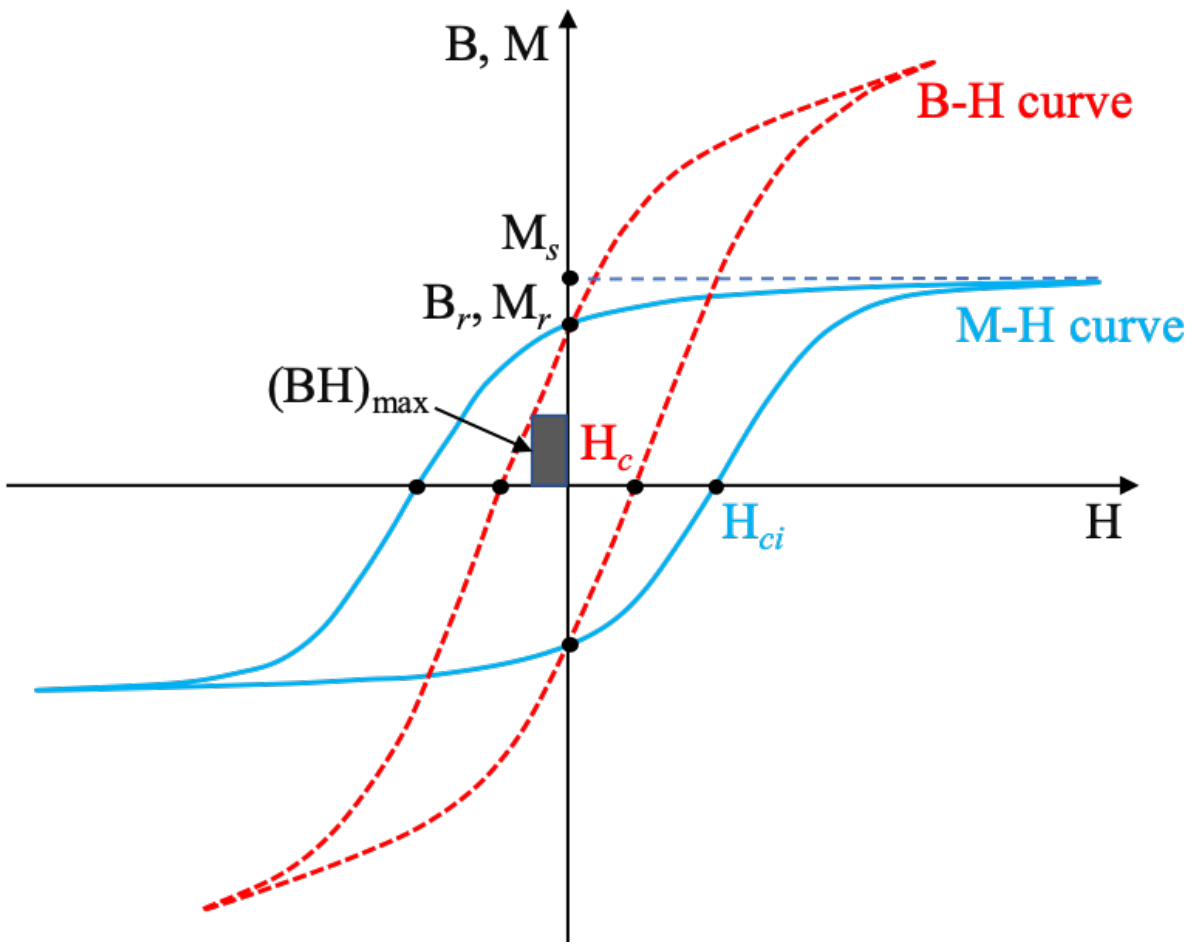

FIG. 1: (color online) Typical hysteresis curves of a FM material. The B-H and M-H loops are denoted by red dashed and blue solid lines, respectively. The black bullets marks the critical values of the key quantities such as the residual induction $B_r$, remanent magnetization $M_r$, coercivity $H_c$, intrinsic coercivity $H_{ci}$ and saturation magnetization $M_s$. The maximal energy product $(BH)_{max}$ is indicated by the shaded region.

Another interesting class of magnetic materials are those with phase transitions (either first-order or second-order) driven by external magnetic fields, leading to the magnetocaloric effect (MCE) and magnetic shape memory effect (MSME). Compared to the 45% efficiency for the best gas-compressing refrigerators, the cooling efficiency of MCE devices based on Gd can reach 60% of the theoretical limit.[32] Moreover, the MCE devices are highly compact and less noisy, giving rise to environmental-friendly solutions for ever-growing demands of cooling on the global scale. Following the thermodynamic Maxwell relation $\frac{\partial S}{\partial B}\big|_T = \frac{\partial M}{\partial T}\big|_B$, optimized MCE can be achieved upon phase transitions with significant changes in the magnetization, which can be easily realized in those compounds with first-order phase transitions (FOPTs). This causes a problem about how to reduce the concomitant hysteresis caused by the athermal nature of FOPTs.[33] On the other hand, the MSME is caused by the domain wall twinning induced by magnetic fields during FOPTs (mostly martensitic transitions), and the corresponding magnetic shape memory alloys (MSMAs) such as Ni-Mn-Ga alloys can be applied as actuators and sensors.[34]

Last but not least, magnetic materials play a pivotal role in the spintronic information technologies.[35] As detailed in Sect. IV G, the first generation of spintronic devices rely on the spin-dependent transport (either diffusive or tunnelling) phenomena which are best represented by the discovery and application of the giant magnetoresistance (GMR) effect and the conjugate spin transfer torque (STT).[36] Whereas the second generation spintronics takes advantage of spin-orbit coupling (SOC) (thus dubbed as spin-orbitronics) and functions via the generation, manipulation, and detection of spin current, engaging both FM and AFM materials.[37] Many materials have been investigated including half-metals (HMs),[38] dilute magnetic semiconductors (DMSs)[39], leading to devices like magnetic random-access memory (MRAM), and spin transistors. Additionally, magnetic materials can also be applied for the storage of information in both analogue and digital forms, where the materials optimization is a trade-off between competing quantities like signal to noise ratio, write-ability with reasonable fields, and long-term stability against thermal fluctuations.[40]

## III. CRITICALITY AND SUSTAINABILITY

Magnetic materials are a prime example where the supply risk of strategic metals, here most importantly rare-earth (RE) elements, might inhibit the future development. In general, resource criticality and sustainability is understood as a concept to assess potentials and risks in using raw materials for certain technologies, particularly strategic metals and their functionalities in emerging technologies.[41] Such principles and evaluation can be transferred to the other material classes subjected to HTP design and future applications. As the application and market for magnetic materials are expected to grow in future technologies, factors such as geological availability, geopolitical situation, economic developments, recyclability, substitutability, ecological impacts, critical competing technologies and the performance of the magnetic materials have to be considered at the beginning in the development of new materials and their constituent elements.

To be specific, the RE elements such as Sm, Dy, and Tb (the latter two are usually used to improve the coercivity and thermal stability of the Nd-Fe-B systems[42]) are of high supply risk due to geopolitical reasons with an expensive price, as specified in the "Critical Materials Strategy" report,[43] leading to a source of concern dubbed as the "rare-earth crisis".[44] Particularly, such RE metals in high demands have relatively low abundance, whereas the abundant light RE elements such as La and Ce may also be utilized to design volume magnets with desired properties. The scenario also applies to transition metal (TM) and main group elements such as Co, Ga, and Ge, which are susceptible to the sustainable availability and can probably be substituted with Mn, Fe, Ni, Al, *etc.* In addition to high prices and low abundance, toxic elements like As and P are another issue, which dictate complex processing to make the resulting compounds useful. That is, not all elements in the periodic table are equally suitable choices for designing materials in practice.

TABLE I: A summary of the main applications of magnetic materials, with representative compounds

| | required properties | materials | applications |
|---|---|---|---|
| permanent magnets | high anisotropy<br>large $M_r$<br>high coercivity $H_c$<br>low permeability<br>high Curie temperature | AlNiCo<br>Ferrite<br>Sm-Co<br>Nd-Fe-B | power generation<br>electric motor<br>robotics |
| soft magnets | low anisotropy<br>large $M_s$<br>high permeability<br>small coercivity<br>low hysteresis<br>low eddy current losses | Fe-Si<br>low-carbon steel<br>Fe-Co | transformer core<br>inductor<br>magnetic field shield |
| MSMA | magnetic phase transition<br>structural reorientation | Ni-Mn-Ga | vibration damper<br>actuator<br>sensor<br>energy harvester |
| magnetocaloric materials | large temperature change $\Delta$T<br>minimal hysteresis<br>mechanical stability | La-Fe-Si<br>Ni-Mn-X<br>$Gd_5(Si,Ge)_4$ | magnetic refrigeration |
| spintronics | strong spin polarization<br>efficient spin injection<br>long spin diffusion length<br>controllable interfaces<br>high ordering temperature | HMs<br>DMS | sensor<br>MRAM<br>spin transistor |
| magnetic storage | medium coercivity<br>large signal-to-noise ratio<br>short writing time $10^{-9}$s<br>long stability time 10 years | Co-Cr<br>FePt | hard disks |

## IV. MAIN CHALLENGES AND POSSIBLE SOLUTIONS

In this section, the fundamental aspects of designing magnetic materials are discussed, with the pending problems and possible solutions illustrated.

### A. New compounds and phase diagram

There have been of the order of $10^5$ inorganic compounds which are experimentally known (*e.g.* from the ICSD database), which amount to a few percent of all possible combinatorial compositions and crystal structures.[45] Therefore, a common task for materials design of any functionality is to screen for unreported compounds by evaluating the stabilities, which can be performed via HTP DFT calculations. Theoretically, the stabilities can be characterized in terms of the following criteria:

1. thermodynamical stability, which can be characterized by the formation energy and distance to the convex hull (Fig. 2), defined as

$$\Delta G = G(\text{target}) - G(\text{comp. phases}) \leq 0, \qquad (1)$$

where $G$(target) and $G$(comp. phases) denote the Gibbs free energy for the target and competing phases, respectively. We note that the Gibbs free energy is a function of temperature T and pressure P, leading to possible metastable phases at finite temperature/pressure. In most cases, only the formation energy with respect to the constituent elements is evaluated, but not the convex hull with respect to the other (either known or unknown) competing phases. It is observed that evaluating the convex hull can reduce the number of predicted stable compounds by one order of magnitude.[46] Thus, it is recommended to carry out the evaluation of convex hull routinely for reasonable predictions, using at least those relevant compounds collected in existing databases as competing phases. Certainly, there are always *unknown* phases which can jeopardize the predictions, but it is believed that the false positive predictions will be significantly reduced with the resulting candidates more accessible for further experimental validation.

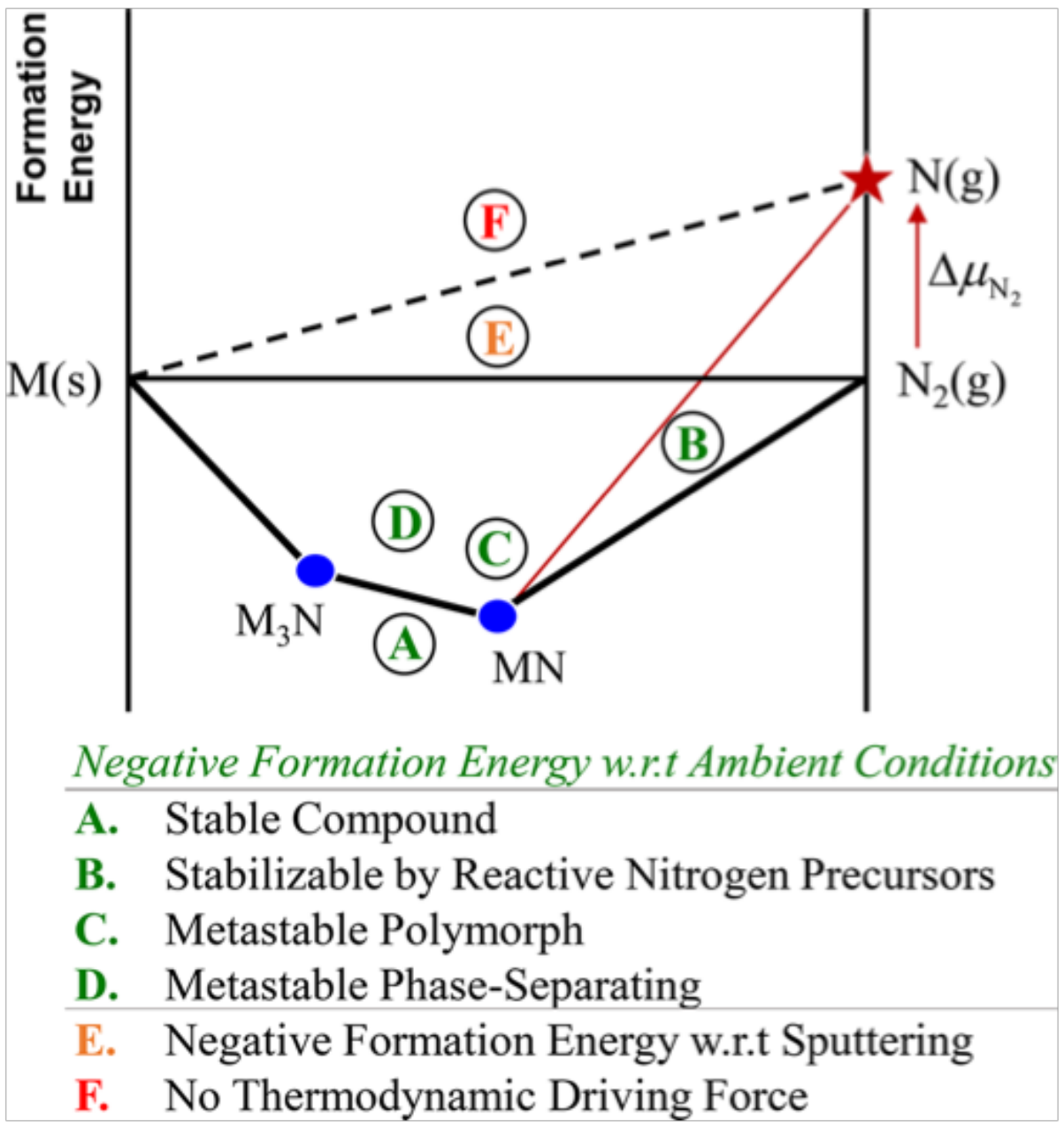

FIG. 2: Sketch of the convex hull for a hypothetical binary metal(M) nitride.[47] Copyright requested.

A few comments are in order. First of all, DFT calculations are usually performed at 0 K and ambient pressure, thus cannot be directly applied to access the metastability at finite temperature or pressure. The pressure can be easily incorporated into calculations using most DFT codes, whereas the temperature constraint can also be remedied by evaluating the Gibbs free energies as discussed later in Sect. IV D. Nevertheless, given that typical solid phase transitions occur around a few hundreds Kelvin which amounts to tens of meV, it is a challenging task to evaluate the phase transition temperature accurately based on DFT calculations. In this regard, the combination of the DFT and CALPHAD[48] methods provides a good solution where experimental measurements can be easily incorporated, in addition to a straightforward generalization to multicomponent systems. The resulting phase diagram will also provide valuable guidance for the experimental synthesis. Furthermore, the stability of metastable phases can be further enhanced by modifying the experimental processes. For instance, precursors can be used in order to reduce the thermodynamic barrier (Fig. 2), *e.g.*, using more reactive nitride precursors like $NH_3$ may allow the synthesis of metastable phases.[47] Lastly, non-equilibrium synthesis techniques such as molecular beam epitaxy (MBE), melt spinning, mechanical alloying, and specific procedure to get nano-structured materials can also be applied to obtain metastable phases, as demonstrated for the $\varepsilon$-phase of MnAl.[49]

Particularly for magnetic compounds, in most cases when evaluating the thermodynamic stability, the formation energies and distances to the convex hull are obtained assuming FM configurations, as done in Materials Project and OQMD. This is justified for the majority of the systems, but we observed that the magnetic states will change the energy landscape drastically for compounds with strong magneto-structural coupling, which will be discussed in detail in Sect. IV C. Another critical problem is how to obtain reliable evaluation of the formation energies for the RE-based intermetallic compounds, where mixing DFT (for the intermetallics) and DFT+U (for the RE elements) calculations are required. This is similar to the case of TM oxides, where additional correction terms are needed to get a reasonable estimation of the formation energies.[50] It is noted that it is a challenging task to do proper DFT+U calculations in a HTP way, where local minima occur very often without good control on the density matrix and additional orbital polarization correction is needed to get correct orbital moments.[51] Therefore, a solution to evaluate the thermodynamic stability is still missing for RE-based intermetallic compounds.

2. mechanical stability, which describes the stability against distortions with respect to small strain. It can be formulated as

$$\Delta E = E^{\text{distorted}} - E^0 = \frac{1}{2}\sum_{i,j} C_{i,j}\epsilon_i\epsilon_j > 0, \qquad (2)$$

where $C_{ij}$ denotes the elastic constant and $\epsilon$ the strain, and $E^0$ denotes the ground state energy from DFT with equilibrium lattice parameters. Depending on the crystalline symmetry, Eq. (2) can be transformed into the generic Born stability conditions,[52] *i.e.*, a set of relationships for the elastic constants which can be straightforwardly evaluated based on DFT.

3. dynamical stability, which describes the stability against atomic displacements due to phonons. In the harmonic approximation,[53] it yields

$$E = E^0 + \underbrace{\frac{1}{2}\sum_{\mathbf{R},\sigma}\sum_{\mathbf{R}',\sigma'} D_{\mathbf{R},\sigma}\Phi^{\sigma,\sigma'}_{\mathbf{R},\mathbf{R}'}D_{\mathbf{R}',\sigma'}}_{>0}, \qquad (3)$$

where $\Phi^{\sigma,\sigma'}_{\mathbf{R},\mathbf{R}'}$ is the force constant matrix, $D_{\mathbf{R},\sigma}$ marks the displacement of atom at $\mathbf{R}$ in the Cartesian direction $\sigma$. The positive definiteness can be assured by diagonalizing the Fourier transformation of $\Phi^{\sigma,\sigma'}_{\mathbf{R},\mathbf{R}'}$, where no negative eigenvalue is allowed. That is, there should be no imaginary phonon mode in the whole Brillouin zone (BZ). On the other hand, if there does exist imaginary phonon modes, particularly at a few specific **q**-points, it suggests that the compounds can probably be stabilized in the correspondingly distorted structure.

Till now, the stabilities are evaluated assuming specific crystal structures. Such calculations can be easily extended to include more structural prototypes based on the recently compiled libraries of prototypes.[54,55] This has been applied successfully to screen for stable $ABO_3$ perovskite[56] and Heusler[57] compounds. One interesting question is whether there are phases with *unknown* crystal structures which might be stable or metastable, giving rise to the question of crystal structure prediction. It is noted that the number of possible structures scales exponentially with respect to the number of atoms within the unit cell, *e.g.*, there are $10^{14}$ ($10^{30}$) structures with 10 (20) atoms per unit cell for a binary compound.[58] There have been well established methods as implemented in USPEX[59] and CALYPSO[60] to predict possible crystal structures. For instance, within the NOVAMAG project, the evolutionary algorithm has been applied to predict novel permanent materials, leading to $Fe_3Ta$ and $Fe_5Ta$ as promising candidates.[61]

To summarize, HTP calculations can be performed to validate the stability of known compounds and to predict possible new compounds. The pending challenges are (a) how to systematically address the stability and metastability, ideally with phase diagrams optimized incorporating existing experimental data, (b) how to extent to the multicomponent cases with chemical disorder and thus entropic free energy, so that the stability of high entropy alloys[62] is accessible, and (c) how to perform proper evaluation of the thermodynamic stability for correlated RE compounds and transition metal oxides.

### B. Correlated nature of magnetism

For magnetic materials, based on (a) how the magnetic moments are formed and (b) the mechanism coupling the magnetic moments, the corresponding theory has a bifurcation into *localized* and *itinerant* pictures.[63] The former dates back to the work of Heisenberg,[64] which applies particularly for strongly correlated TM insulators and RE-based materials with well-localized $f$-electrons.The local magnetic moments can be obtained following the Hund's rule via

$$\mu_{\text{eff}} = g_J \sqrt{J(J+1)} \mu_B, \tag{4}$$

where $g_J$ is the Landé g-factor, $J$ denotes the total angular momentum, and $\mu_B$ is the Bohr magneton. The magnetic susceptibility above the critical ordering temperature (*i.e.*, in the paramagnetic (PM) states) obeys the Curie-Weiss law:[65]

$$\chi(T) = \frac{C}{T - \theta_W}, \tag{5}$$

where $C$ is a material-specific constant, and $\theta_W$ denotes the Curie-Weiss temperature where $\chi$ shows a singular behavior. The Weiss temperature $\theta_W$ can have positive and negative values, depending on the resulting ferromagnetic (FM) and antiferromagnetic (AFM) ordering, whereas its absolute value is comparable to the Curie/Néel temperature. The Weiss temperature corresponds to the *molecular field* as introduced by Weiss,[66] which can be obtained via a mean-field approximation of the general Heisenberg Hamiltonian:

$$H = -\frac{1}{2} \sum_{i,j} J_{ij} \mathbf{S}_i \cdot \mathbf{S}_j, \tag{6}$$

where $J_{ij}$ denotes the interatomic exchange interaction. Specifically, $\mathbf{S}_i \cdot \mathbf{S}_j$ is approximated to $S_i^z \langle S_j^z \rangle + \langle S_i^z \rangle S_j^z$ by neglecting the fluctuations and hence the spin flip term $S_i^+ S_j^-$, leading to an effective field proportional to the magnetization corresponding to the exchange field in spin-polarized DFT.[67]

On the other hand, for intermetallic compounds with mobile conduction electrons, the magnetic ordering is caused by the competition between the kinetic energy and magnetization energy. Introducing the exchange energy $I = J/N$ where $J$ is the averaged exchange integral originated from the *intra-atomic inter-orbital* Coulomb interaction and N the number of atoms in the crystal, a ferromagnetic state can be realized if[68]

$$I\nu(E_F) > 1, \tag{7}$$

where $\nu(E_F)$ is the density of states (DOS) at the Fermi energy $E_F$ in the nonmagnetic state. The so-obtained itinerant picture has been successfully applied to understand the occurrence of ferromagnetism in Fe, Co, Ni, and many intermetallic compounds including such elements.[69]

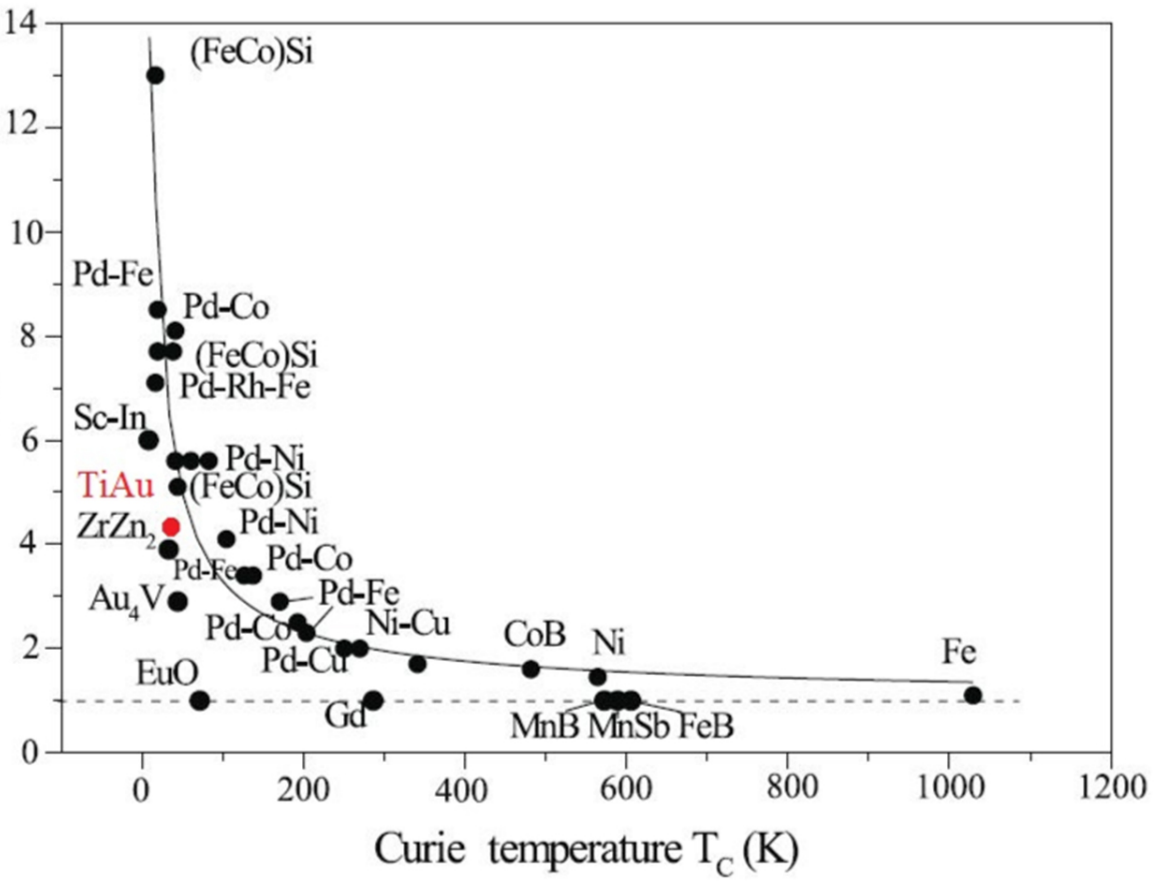

FIG. 3: The Rhodes-Wohlfarth ratio $q_c/q_s$ for various magnetic materials.[65]Copyright requested

However, the real materials in general cannot be classified as either purely localized or itinerant and are mostly at a crossovers between two limits. As shown in Fig. 3, where the ratio of Curie-Weiss constant $q_c$ and the saturation magnetization $q_s$ is plotted with respect to the Curie temperature, the $q_c/q_s$ ratio should be equal to 1 for localized moments as indicated by EuO, whereas for the itinerant moments $q_c/q_s \gg 1$. It is interesting that the $q_c/q_s$ ratio for Ni is close to 1 while it is a well known itinerant magnet. More interestingly, there exists a class of materials such as TiAu and $ZrZn_2$ where there is no partially filled $d$- or $f$-shells but they are still displaying ferromagnetic behavior.[65]

It is noted that the localized moment picture based on the Heisenberg model has limited applicability on the intermetallic magnets, while the Stoner picture fails to quantitatively describe the finite temperature magnetism, *e.g.*, it would significantly overestimate the Curie temperature and leads to vanishing moment and hence no Curie-Weiss behavior above $T_C$. A universal picture can be developed based on the Hubbard model,[70]

$$H = \sum_{ij\sigma} t_{ij} a^{\dagger}_{i\sigma} a_{j\sigma} + \sum_{i} U n_{i\uparrow} n_{i\downarrow} \tag{8}$$

where $t_{ij}$ is the hopping parameter between different sites, $n$ is the number of electrons, and U denotes the on-site Coulomb interaction. As a matter of fact, the Stoner model is a mean field approximation of the Hubbard model in the weakly correlated limit andthe Heisenberg model can be derived in the strongly coupled limit at half-filling.[71] Another unified theory has been formulated by Moriya considering self-consistent renormalization of the spin fluctuations in the static and long-wave length limits,[63] which works remarkably well for the weak itinerant magnets such as $ZrZn_2$.[65] In terms of the Hubbard model, which can be further casted into the DFT + dynamical mean-field theory (DMFT) framework,[72] where the hoppings are obtained at the DFT level and the onsite electron-electron correlations are evaluated accurately locally including *all* orders of Feynman diagrams. Recently, the DFT+DMFT methods have been applied on $ZrZn_2$,[73] revealing a Fermi liquid (FL) behavior of the Zr-4$d$ electrons which are responsible for the formation of magnetic moments. In this regard, it provides a universal solution which can be applied for magnetic materials from localized to itinerant limits, which is better than bare DFT. Importantly, for correlated TM oxides and RE compounds, the $d$- and $f$-moments are mostly localized, leading to narrow bands where the local quantum fluctuations are significant and hence the spin fluctuations,[74] which are naturally included in DFT+DMFT. Nevertheless, the usually performed single-site DFT+DMFT calculations cannot capture the transverse magnetic excitations particularly the long wave-length excitations critical for the behavior around $T_C$. Detailed discussions will be presented in Sect. IV D. We noted that the classification of itinerant and localized magnetism is not absolute, *e.g.*, the Ce-4$f$ shell changes its nature depending on the crystalline environment.[75]

In this sense, DFT as a mean field theory has been successfully applied to magnets of the itinerant and localized nature. In the latter case, the DFT+U[76] method is usually applied to account for the strong electronic correlations. However, there are known cases where DFT fails, *e.g.*, it predicts a FM state for FeAl while experimentally the system is paramagnetic.[77] Also, the local density approximation (LDA) gives wrong ground state of Fe, *i.e.*, LDA predicts the nonmagnetic fcc phase being more stable than the FM bcc phase.[78] Therefore, cautions are required when performing DFT calculations on magnetic materials and validation with experiments is always called for. On the other hand, the DFT+DMFT method is a valuable solution which covers the whole range of electronic correlations, but the key problem is that the current implementations do not allow automated set-up and efficient computation on a huge number of compounds.

### C. Magnetic ordering and ground states

Assuming the local magnetic moments are well defined, the Heisenberg Hamiltonian Eq. (IV B) is valid to describe the low temperature behavior, the interatomic exchange $J_{ij}$ is the key to understand the formation of various long-range ordered states.There are three types of mechanisms, based on the distance between the magnetic moments and how do they talk to each other.

1. *direct exchange* for atoms close enough so that the wave functions have sufficient overlap. Assuming two atoms each with one unpaired electron, the Coulomb potential is reduced for two electrons localized between the atoms when the atoms are very close to each other, leading to AFM coupling based on the Pauli's exclusion principle. On the other hand, when the distance between two atoms becomes larger, the electrons tend to stay separated from each other by gaining kinetic energy, resulting in FM coupling. Thus, the direct exchange is short ranged, which is dominant for the coupling between the first nearest neighbors. This leads to the Bethe-Slater curve, where the critical ratio between the atomic distance and the spatial extent of the 3$d$-orbital is about 1.5 which separate the AFM and FM coupling.[79]

2. *indirect exchange* mediated by the conduction electrons for atoms without direct overlap of the wave functions. It can be understood based on the Friedel oscillation, where the conduction electrons try to screen a local moment, thus forming long range oscillating FM/AFM exchange coupling, leading to the RKKY-interaction named after Ruderman, Kittel, Kasuya, and Yoshida.[80] It is the dominant exchange coupling for atoms beyond the nearest neighbour in TM magnets and localized 4$f$-moments (mediated by the $sp$-electrons) in RE magnets.

3. *superexchange* for insulating compounds with localized moments, which is driven by kinetic energy gain via the virtual excitations of local spin between magnetic ions through the bridging nonmagnetic elements, such as oxygen in TM oxides. The superexchange can favor both AFM and FM couplings depending on the orbital occupation and the local geometry, which has been formulated as the Goodenough-Kanamori-Anderson rules.[81]

We note that in real materials, more than one type of exchange coupling should be considered, *e.g.*, the direct exchange and RKKY exchange for intermetallic compounds, and the competition of direct exchange and superexchange in TM oxides.

Based on the discussions above, the exchange coupling between atomic moments can be either FM or AFM, depending on the active orbitals, distances, and geometries. Such AFM/FM exchange couplings induce various possible magnetic configurations, which give rise to a complication for HTP screening of magnetic materials. That is, the total energy difference for the same compounds with different magnetic states can be significant. This can be roughly estimated by the Curie temperature ($T_C$), i.e., about 0.1 eV corresponding to the $T_C$ (1043 K) of bcc Fe. Thus, magnetic ground state is important not only for the electron structure but also for the thermodynamic stability. However, the number of possible AFM configurations which should be considered to define the magnetic ground state can be big, depending on the crystal structures and the magnetic ions involved. Therefore, identifying the magnetic ground state is the most urgent problem to be solved for predictive HTP screening on magnetic materials.

There have been several attempts trying to develop a solution. The most straightforward way is to collect all possible magnetic configurations from the literature for a specific class of materials or consider a number of most probable states, such as done for the half-Heuslers.[83] However, it is hard to be exhaustive and more importantly it is only applicable for one structural prototype. Horton *et al.* developed a method to enumerate possible magnetic configurations followed by HTP evaluation of the corresponding total energies.[84] It is mostly applicable to identify the FM ground state, as the success rate is about 60% for 64 selected oxides, which might due to the correlated nature of such materials. Another interesting method is the firefly algorithm, as demonstrated for $NiF_2$ and $Mn_3Pt$,[85] where the magnetic configurations are confined within the primitive cell. A systematic way to tackle the problem is to make use of maximal magnetic subgroups,[82] where the magnetic configurations are generated in a progressive way taking the propagation vectors as a control parameter. We have implemented the algorithm and applied it successfully on the binary intermetallic compounds with the $Cu_3Au$-type structure. It is observed that the landscape of convex hull changes significantly after considering the magnetic ground state, as shown in Fig. 4 for the binary Mn-Ir systems. That is, the magnetic ground state matters not only for the electronic structure but also for the thermodynamic stability of magnetic materials, confirming our speculation in Sect. IV A. Recently, the genetic algorithm has been implemented to generate magnetic configurations and combined with DFT calculations the energies for such configurations can be obtained in order to search for the magnetic ground state.[86] It has been applied successfully on FeSe, $CrI_3$ monolayers, and $UO_2$, where not only the collinear but also noncollinear configurations can be generated, which is also interesting for future exploration.

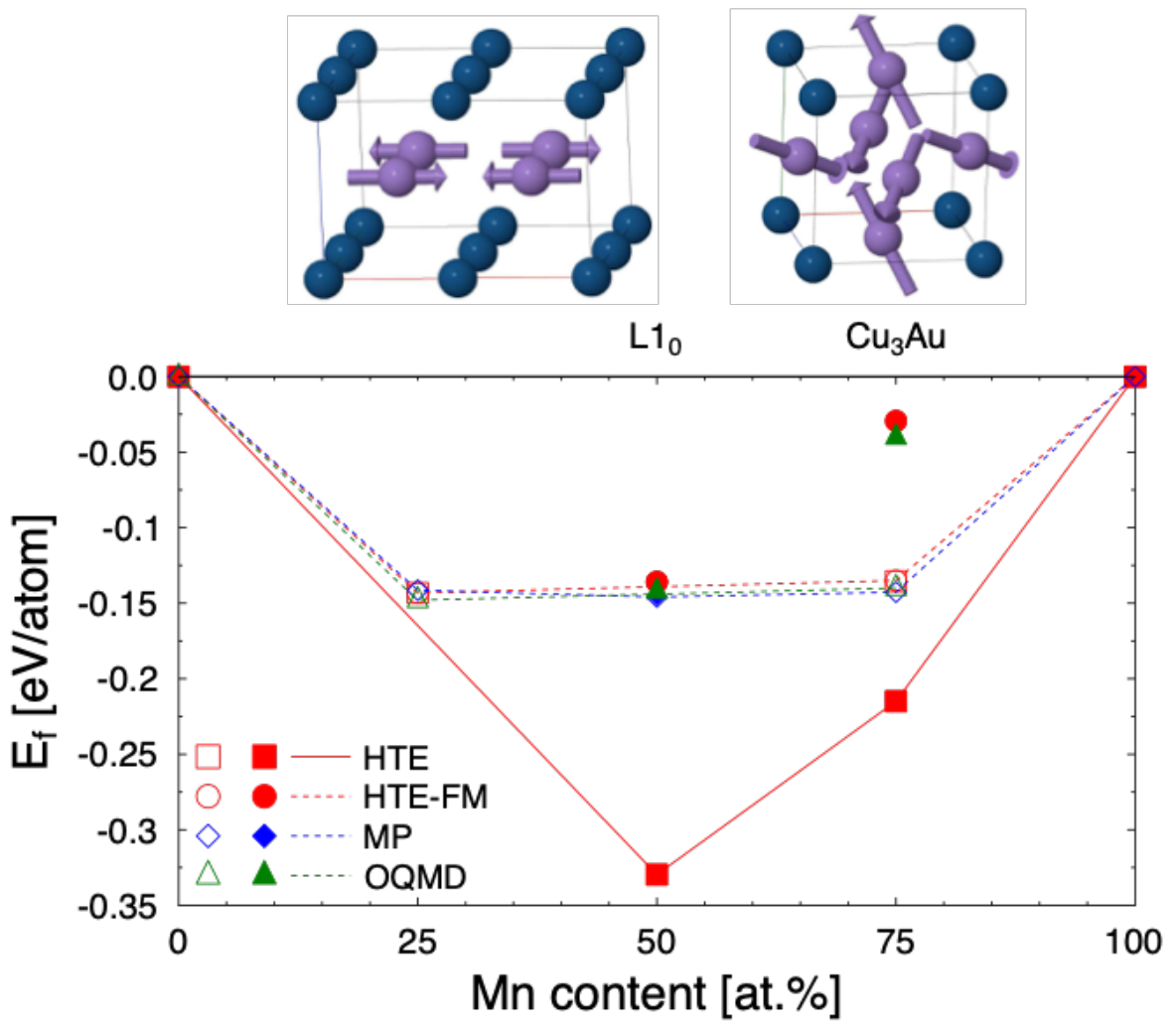

FIG. 4: The binary convex hull for Mn-Ir. Dashed (solid) lines denotes the convex hull obtained assuming ferromagnetic configurations (with real magnetic ground states obtained via systematic calculations by maximal magnetic subgroups). The magnetic ground states for MnIr and $Mn_3Ir$ are display on top of the figure.[82]Copyright requested

Furthermore, the interatomic exchange $J_{ij}$ in Eq. (IV B) can be evaluated based on DFT calculations, which can then be used to find out the magnetic ground states via Monte Carlo modelling. There are different methods on evaluating the $J_{ij}$. The most straightforward one is the energy mapping, where the $J_{ij}$ for each pair of moments are obtained by the difference of total energies for the FM and AFM configurations.[87] Such a method can be traced back to the broken symmetry method firstly proposed by Noodleman[88] and later generalized by Yamaguchi,[89] where one of the spin state used in energy mapping is not the eigenstate of the Heisenberg Hamiltonian (Eq. IV B). In this regard, for magnetic materials with multiple magnetic sublattices, the isotropic $J_{ij}$ can be obtained by performing least square fitting of the DFT total energies of a finite number of imposed magnetic configurations, *e.g.*, using four-state mapping method.[87,90]

Based on a two-site Hubbard model, it is demonstrated that the energy mapping method is only accurate in the strong coupling limit, *i.e.*, insulating states with well defined local moments.[91] Additional possible problems for such an approach are (a) the supercell can be large in order to get the long-range interatomic exchange parameters and (b) the magnitude of local moments might change for the FM and AFM calculations resulting in unwanted contribution from the longitudinal excitations. Thus, this method is most applicable for systems with well-defined local moments. In this case, the $J_{ij}$ can also be evaluated based on the perturbation theory, which can be done with the help of Wannier functions hence there is no need to generate supercells.[92] The artificial longitudinal excitations can also be suppressed by performing constrained DFT calculations.[93]

There are two more systematic ways to evaluate the $J_{ij}$. One is based on the so-called frozen magnon method,[94] where the total energies for systems with imposed spin-waves of various $\mathbf{q}$ vectors are calculated and afterwards a back Fourier transformation is carried out to parameterize the real space $J_{ij}$.[95] This method demands the implementation of noncollinear magnetism in the DFT codes. Another one is based on the magnetic force theorem,[96] where the $J_{ij}$ is formulated as a linear response function which can be evaluated at both the DFT and DFT+DMFT levels.[97] It is noted that the effective spin-spin interaction is actually a second order tensor,[98] which yields

$$\frac{1}{2}\sum_{i,j}\mathbf{S}_i\cdot\mathbf{J}_{ij}\cdot\mathbf{S}_j = \frac{1}{2}\sum_{i,j}\left[\frac{1}{3}\mathrm{Tr}(\mathbf{J}_{ij})\mathbf{S}_i\cdot\mathbf{S}_j + \mathbf{S}_i(\frac{1}{2}(\mathbf{J}_{ij}+\mathbf{J}^t_{ij})-\mathbf{J}_{ij})\mathbf{S}_j + \frac{1}{2}(\mathbf{J}_{ij}-\mathbf{J}^t_{ij})\mathbf{S}_i\times\mathbf{S}_j\right] \tag{9}$$

where the Heisenberg term in Eq. (IV B) corresponds to the isotropic exchange $\frac{1}{3}\mathrm{Tr}(\mathbf{J}_{ij})$, the symmetric traceless part $\mathbf{J}^{\mathrm{sym}}_{ij} = \frac{1}{2}(\mathbf{J}_{ij}+\mathbf{J}^t_{ij})-\mathbf{J}_{ij}$ is usually referred as anisotropic exchange, and the antisymmetric part $\mathbf{J}^{\mathrm{antisym}}_{ij} = \frac{1}{2}(\mathbf{J}_{ij}-\mathbf{J}^t_{ij})$ represents the Dzyaloshinsky-Moriya interaction (DMI). It is noted that DMI is the crucial parameter to form Skyrmions.[24] Also, in order to address complex magnetic orderings, *e.g.*, in 2D magnetic materials, the full tensor should be evaluated consistently to construct fully-fledged spin models.

In short, the essential challenges are (a) how to obtain the magnetic ground states in a HTP manner and (b) how to evaluate the exchange parameters. As we discussed, there are a few possible solutions for the former question. Regarding the latter, as far as we are aware, there has been no reliable implementation where the exchange parameters can be evaluated in an automated way, where either the DFT part of the codes needs careful adaption for materials with diverse crystal structures or only specific components of the exchange parameters can be evaluated.[99]

### D. Magnetic fluctuations

The discussions till now have been focusing on the magnetic properties at zero Kelvin, however, the thermodynamic properties and the magnetic excitations of various magnetic materials are also fundamental problems which should be addressed properly based on the microscopic theory. To this goal, the magnetic fluctuations driven by temperature shall be evaluated, which will destroy the long-range magnetic ordering at $T_C/T_N$, resulting in a paramagnetic state with fluctuating moments. The working horse DFT can in principle be generalized to finite temperature,[100] but to the best of our knowledge there is no implementation into the DFT codes on the market. Specifically for the magnetic excitations, the *ab initio* spin dynamics formalism proposed by Antropov *et al.*[101] is accurate but it is demanding to get implemented, thus it has only been applied to simple systems such as Fe. In the following, we will focus on (a) the atomistic spin models, (b) the disordered local moment approximation, and (c) the DFT+DMFT methods, with the thermodynamic properties at finite temperature in mind.

From the physics point of view, there are two types of excitations, *i.e.*, transversal and longitudinal excitations. For itinerant magnets, the former refers to the collective fluctuations of the magnetization directions which dominates at the low-temperature regime, whereas the latter is about the spin-flip excitations (*i.e.*, Stoner excitations) leading to the reduction of magnetic moments. Correspondingly for the localized moments, the occupation probability of the atomic multiplets varies with respect to temperature, giving rise to the temperature dependent magnitude of local moments, while the interatomic exchange parameters of the RKKY type will cause collective transversal spin waves as well.

One of the most essential intrinsic magnetic properties is the magnetic ordering temperature, *i.e.*, $T_C$ ($T_N$) for FM (AFM) materials. The theoretical evaluation of the $T_C/T_N$ usually starts with the Heisenberg model (Eq. (IV B)) with the $J_{ij}$ parameters obtained from DFT calculations, as detailed in Sect. IV C. As the spin wave precession energy is smaller than the band width and the exchange splitting, it is justified to neglect the precession of magnetization due to the spin waves when evaluating the electronic energies, dubbed as the adiabatic approximation, the critical temperature can be obtained by statistical averaging starting from the Heisenberg model. Both mean-field approximation and ran-dom phase approximation can be applied, where the former fails to describe the low-temperature excitations.[102] In this way, the longitudinal excitations with comparable energies[103] can be treated based on parameterized models.[95,104] Classical Monte Carlo is often used, leading to finite specific heat at zero Kelvin,[105] while quantum Monte Carlo suffers from the sign problem when long-range exchange parameters with alternative signs are considered. One solution to this problem is to introduce the quantum thermostat.[106] It is noted that more terms such as external magnetic fields and magnetic anisotropy can be included, leading to atomistic spin dynamics which is valuable to develop a multi-scale understanding of magnetic properties.[107,108]

Another fundamental quantity is the Gibbs free energy which is crucial to elucidate the magneto-structural phase transitions. In general, the Gibbs free energy can be formulated as[109]

$$G(T,P,\mathbf{e_m}) = H(T,V,\mathbf{e_m}) + PV = E(V,\mathbf{e_m}) + F^{\text{electronic}}(V,T) + F^{\text{lattice}}(V,T) + F^{\text{magnetic}}(V,T) + PV, \tag{10}$$

where $\mathbf{e_m}$ denotes the magnetization direction, T, P, and V refer to the temperature, pressure, and volume, respectively. The electronic, lattice, and magnetic contributions to the free energy can be considered as independent of each other, because the typical time scale of electronic, lattice, and magnetic dynamics is about $10^{-15}$s, $10^{-12}$s, and $10^{-13}$s, respectively. This approximation is not justified anymore for the paramagnetic states where the spin decoherence time is about $10^{-14}$s,[110] which will be discussed in detail below.

To evaluate $F^{\text{magnetic}}$, the disordered local moment (DLM) method[111] provides a valuable solution for both magnetically ordered and disordered (e.g., paramagnetic) states. The DLM method also adopts the adiabatic approximation, and it includes two conceptual steps:[112] (1) the evolution of the local moment orientations and (2) the electronic structure corresponding to each specific configuration. A huge number of orientational configurations are required in order to accurately evaluate the partition function and hence the thermodynamic potential, which can be conveniently realized using the multiple scattering theory formalism based on the Green's functions, namely the KKR method.[113,114] The recent generalization[115] into the relativistic limit enables accurate evaluation of the temperature-dependent MAE for both TM-based[115] and $4f$-$3d$ magnets,[116] and also the magnetic free energies with even frustrated magnetic structures.[112] Nevertheless, the codes with the KKR method implemented requires sophisticated tuning for complex crystalline geometries and the evaluation of forces is not straightforward.

The DLM method can also be performed using supercells by generating magnetic configurations like special quasi-random structures (SQS), which is invented to model the chemical disorder.[117,118] It has been applied to simulate the paramagnetic states with supercells of equal number of up and down moments, particularly to evaluate the lattice dynamics as the forces can be obtained in a straightforward way. Moreover, configurational averaging should be done in order to model local defects, which can be achieved by generating a number of supercells with randomly distributed moments which sum up to zero.[118] Spin-space averaging can be performed on top to interpolate between the magnetically ordered and the paramagnetic states, as done for bcc Fe.[119] In this way, the spin-lattice dynamics can be performed to obtain the thermodynamic properties for the paramagnetic states,[120] which cannot be decoupled as discussed above. It is noted that usually the spin-lattice dynamics is performed based on model parameters,[121] where the instantaneous electronic response is not considered. The DLM method with supercell has also been recently combined with the atomistic spin dynamics with in general noncollinear spin configurations.[122] It is noted that the local moments should be very robust in order to perform such calculations, otherwise significant contribution from longitudinal fluctuations might cause trouble.

Another promising method to address all the issues discussed above is the DFT+DMFT methods. The DMFT method works by mapping the lattice Hubbard model onto the single-site Anderson impurity model, where the crystalline environment acts on the impurity via the hybridization function.[72] Continuous time quantum Monte Carlo has been considered to be the state-of-the-art impurity solver.[123] Very recently, both the total energies[124] and forces[125] can be evaluated accurately by performing charge self-consistent DFT+DMFT calculations, with a great potential addressing the paramagnetic states. Nevertheless, the local Coulomb parameters are still approximated, though there is a possibility to evaluate such quantities based on constrained RPA calculations.[126] One additional problem is that the mostly applied single-site DFT+DMFT does not consider the intersite magnetic exchange, which will overestimate the magnetic critical temperature. We suspect that maybe atomistic modelling of the long wavelength spin waves will cure the problem, instead of using expensive cluster DMFT.

Overall, the key challenge to obtain the thermodynamic properties of magnetic materials is the accurate evaluation of the total Gibbs free energy including longitudinal and transversal fluctuations, together with coupling to the lattice degree of freedom. DFT+DMFT is again a decent solution despite known constraints as discussed above, but there is no proper implementation yet ready for HTP calculations.

### E. Magnetic anisotropy and permanent magnets

#### 1. Origin of magnetocrystalline anisotropy

As another essential intrinsic magnetic property, MAE can be mostly attributed to the SOC,[127] which couples the spontaneous magnetization direction to the underlying crystal structure, *i.e.*, breaks the continuous symmetry of magnetization by developing anisotropic energy surfaces.[128] Like spin, SOC is originated from the relativistic effects,[129] which can be formulated as (in the Pauli two-component formalism):

$$H_{\mathrm{SOC}} = -\frac{\nabla V}{2m^2c^2}\mathbf{s}\cdot\mathbf{l} = \xi\mathbf{s}\cdot\mathbf{l}, \tag{11}$$

where $\nabla V$ marks the derivative of a scalar electrostatic potential, $m$ the mass of electrons, $c$ the speed of light, $\mathbf{s}$ ($\mathbf{l}$) the spin (orbital) angular momentum operator. $\xi$ indicates the magnitude of SOC, which is defined for each $l$-shell of an atom, except that for the $s$-orbitals where $l = 0$ leading to zero SOC. The strength of atomic SOC can be estimated by $\xi \approx \frac{2}{5}(\epsilon_{d_{5/2}} - \epsilon_{d_{3/2}})$ for the d-orbitals and $\xi \approx \frac{2}{7}(\epsilon_{d_{7/2}} - \epsilon_{d_{5/2}})$ for the $f$-orbitals, respectively.[130] This leads to the average magnitude of $\xi$ for the $3d$-orbitals of $3d$ TM atoms is about 60 meV (*e.g*, from 10 meV (Sc) to 110 meV (Cu)[131]), while that of the $4f$-orbitals for lanthanide elements can be as large as 0.4 eV. Therefore, the larger the atomic number is, the larger the magnitude of $\xi$, *e.g.*, enhanced SOC in $4d$- and $5d$-elements induces many fascinating phenomena.[132] The electrostatic potential $V$ can be modulated significantly at the surfaces or interfaces, leading to the Rashba effect which is interesting for spintronics.[133]

Conceptually, the MAE can be understood in such a way that SOC tries to recover the orbital moment which is quenched in solids, *i.e.*, it is attributed to the interplay of exchange splitting, crystal fields, and SOC.[134] Based on the perturbation theory, the MAE driven by SOC can be expressed as:[135]

$$\begin{aligned}\mathrm{MAE} = &\underbrace{-\frac{1}{4}\xi\mathbf{e_m}\cdot[\delta\langle\mathbf{l}^{\downarrow}\rangle - \delta\langle\mathbf{l}^{\uparrow}\rangle]}_{\text{orbital moment}} \\ &+ \underbrace{\frac{\xi^2}{\Delta E_{ex}}[10.5\mathbf{e_m}\cdot\delta\langle\mathbf{T}\rangle + 2\delta\langle(l_\zeta s_\zeta)^2\rangle]}_{\text{spin flip}},\end{aligned} \tag{12}$$

where $\mathbf{e_m}$ stands for the unit vector along the magnetization direction. $\mathbf{T} = \mathbf{e_m} - 3\vec{\mathbf{r}}(\vec{\mathbf{r}}\cdot\mathbf{e_m}) \approx -\frac{2}{7}\mathbf{Q}\cdot\mathbf{e_m}$ denotes the anisotropic spin distribution, with $\vec{\mathbf{r}}$ indicating the position unit vector, $\mathbf{Q} = \mathbf{l}^2 - \frac{1}{3}l^2$ indicates the charge quadrupole moment. It is noted that the spin-flip contribution is of higher order with respect to $\xi$, which is significant for compounds with significant SOC. On the other hand, for the orbital momentum term, it will be reduced to $\mathrm{MAE} = -\frac{1}{4}\xi\mathbf{e_m}\cdot\delta\langle\mathbf{l}^{\downarrow}\rangle$ for strong magnets with the majority channel fully occupied,[136] suggesting that the magnetization prefers to be aligned along the direction where the orbital moment is of larger magnitude.

Following the perturbation theory, the MAE can be enhanced by engineering the local symmetry of the magnetic ions. It is well understood that the orbital moments are originated from the degeneracy of the atomic orbitals which are coupled by SOC.[135] Therefore, when the crystal fields on the magnetic ions lead to the degeneracy within the $\{\mathrm{d}_{xy}, \mathrm{d}_{x^2+y^2}\}$ or $\{\mathrm{d}_{yz}, \mathrm{d}_{xz}\}$ orbitals, the MAE can be significantly enhanced, depending also on the orbital occupations. This suggests, local crystalline environments of the linear, trigonal, trigonal bipyramid, pentagonal bipyramid, and tricapped trigonal prism types are ideal to host a possible large MAE, because the resulting degeneracy in both $\{\mathrm{d}_{xy}, \mathrm{d}_{x^2+y^2}\}$ and $\{\mathrm{d}_{yz}, \mathrm{d}_{xz}\}$ orbital pairs.[137] For instance, Fe atoms in $Li_2Li_{1-x}Fe_xN$ located on linear chains behave like RE elements with giant anisotropy,[138] and Fe monolayers on InN substrates exhibit a giant MAE as large as 54 meV/u.c. due to the underlying trigonal symmetry.[139] This leads to an effective way to tailor the MAE by adjusting the local crystalline geometries, as demonstrated recently for magnetic molecules.[140]

Technically, MAE is evaluated as the difference of total energies between two different magnetization directions ($\mathbf{e}^1_\mathbf{m}$ and $\mathbf{e}^2_\mathbf{m}$):

$$\begin{aligned}\mathrm{MAE} &= E_{\mathbf{e}^1_\mathbf{m}} - E_{\mathbf{e}^2_\mathbf{m}} \\ &\approx \sum_{\mathbf{k},i=1}^{\text{occ.}} \varepsilon_i(\mathbf{k}, \mathbf{e}^1_\mathbf{m}) - \sum_{\mathbf{k},i=1}^{\text{occ.}} \varepsilon_i(\mathbf{k}, \mathbf{e}^2_\mathbf{m})\end{aligned} \tag{13}$$

In general, MAE is a small quantity with a typical magnitude between $\mu$eV and meV, as a difference of two numbers which are orders of magnitude larger. In this regard, it is very sensitive to the numerical details such as exchange-correlation functionals, $k$-point convergence, implementation of SOC, and so on, leading to expensive full relativistic calculations particularly for compounds with more than 10 atoms per unit cell. To get a good estimation, force theorem is widely used (second line of Eq. (IV E 1)), where MAE is approximated as the difference of the sum of DFT energies for occupied states obtained by one-step SOC calculations.[141] We want to emphasize that the $k$-point integration over the irreducible part of the Brillouin zone defined by the Shubnikov group of the system is usually required, suggesting (a) symmetry should be applied with caution, *i.e.*, the actual symmetry depends on the magnetization direction and (b) there is no hot zone with dominant contribution.[142] In this regard, the recently developed Wannier interpola-tion technique may provide a promising solution to the numerical problem.[143,144]

### 2. *Permanent magnets: Rare-earth or not?*

From the materials point of view, there are two classes of widely used permanent magnets, namely, the TM-based ferrite and AlNiCo, and the high performance RE-based $Nd_2Fe_{14}B$ and $SmCo_5$. One crucial factor to distinguish such two classes of materials is whether the RE elements are included, which causes significant MAE due to the competition of enhanced SOC and reduced band width for the $4f$-orbitals.[131] Hereafter these two classes of permanent magnets will be referred as RE-free and $4f$-$3d$ magnets. Given the fact that the research on these four systems is relatively mature, the current main interest is to identify the so-called *gap magnets*,[145] *i.e.*, material systems with performances lying between these two classes of permanent magnets.

Since MAE is originated from SOC, it is an inert atomic property, *i.e.*, its strength cannot be easily tuned by applying external forces which are negligible comparing to the gradient of the electrostatic potential around the nuclei. In addition, the exchange splitting is mostly determined by the Coulomb interaction.[69] Thus, the knob to tailor MAE is the hybridization of the atomic wave functions with those of the neighboring atoms, which can be wrapped up as crystal fields. On the one hand, for the $4f$-$3d$ magnets, the MAE is mainly originated from the non-spherical distribution of the $4f$-electrons,[146] which can be expressed based on the local crystal fields on the $4f$-ions.[147] On the other hand, for TM-based magnetic materials, the orbital degree of freedom can be engineered to enhance the MAE by manipulating the crystal fields via fine tuning the crystalline symmetries, as discussed above following the perturbation theory. For instance, the MAE of Co atoms can be enhanced by three orders of magnitude, from 0.001 meV in fcc Co to 0.06 meV/atom in hcp Co driven by the reduction of crystalline symmetry,[148] and further to about 60 meV for Co adatoms on MgO[149] or Co dimers on Benzene attributed to the energy levels strongly affected by the crystal fields.[150] Such a concept can also be applied to improve the MAE of bulk magnetic materials, such as in $Li_2Li_{1-x}Fe_xN$[138] and FeCo alloys,[151] where the $d$-orbitals strongly coupled by SOC (such as $d_{xy}$ and $d_{x^2-y^2}$ orbitals) are adjusted to be degenerate and half-occupied by modifying the local crystalline geometry and doping.

For the RE-free magnets, the theoretical upper limit[152] of MAE as high as 3.6 MJ/m$^3$. That is, there is a strong hope that RE-free materials can beengineered for high performance permanent magnets. There have also been a number of compounds investigated following this line, focusing on the Mn-, Fe-, and Co-based compounds.[153] The Mn-based systems are particularly interesting, as Mn has the largest atomic moments of 5 $\mu_B$ which will be reduced in solid materials due to the hybridization. However, the interatomic exchange between Mn moments has a strong dependence on the interatomic distance, *e.g.*, the Mn moments tend to couple antiferromagnetically (ferromagnetically) if the distance is about 2.5-2.8 Å(>2.9Å).[29] Nevertheless, there are a few systems such as MnAl[154], MnBi[155], and Mn-Ga[156] showing promising magnetic properties to be optimized for permanent magnet applications. On the other hand, for the Fe-based systems, FePt[157] and FeNi[158] have been extensively investigated. One key problem for these two compounds is that there exists strong chemical disorder with depends on the processing. For instance, the order-disorder transition temperature for FeNi is about 593 K, which demands extra long time of annealing to reach the ordered phase with strong anisotropy.[159]

Given that the $M_s$ and MAE can be obtained in a straightforward way for most TM-based magnetic materials, the main strategy for searching novel RE-free permanent magnets is to screen over a number of possible chemical compositions and crystal structures. One particularly interesting direction is to characterize the metastable phases, such as $Fe_{16}N_2$[160,161] and MnAl.[49] Such phases can be obtained via non-equilibrium synthesis, such as MBE, melt spinning and mechanical alloying. In this sense, the permanent magnets of thin films is an oxymoron, *i.e.*, they can provide valuable information about how to engineer the intrinsic properties but it is challenging to obtain bulk magnets. It is also noted that pure DFT might not be enough to account for the magnetism in tricky cases due to the missing spin fluctuations as demonstrated in $(FeCo)_2B$.[162]

For the $4f$-$3d$ compounds, it is fair to say that there is still no consistent quantitative theory at the DFT level to evaluate their intrinsic properties, though both $SmCo_5$ and $Nd_2Fe_{14}B$ were discovered decades ago. The most salient feature common to the $4f$-$3d$ magnetic compounds is the energy hierarchy, driven by the coupling between RE-$4f$ and TM-$3d$ sublattices.[163] First of all, the $T_C$ is ensured by the strong TM-TM exchange coupling. Furthermore, the MAE is originated from the interplay of RE and TM sublattices. For instance, the Y-based compounds usually exhibit significant MAE, such as YCo5,[164] whereas for the RE atoms with nonzero $4f$ moments, the MAE is significantly enhanced due to non-spherical nature of the RE-$4f$ charges[146] caused by the crystal fields.[147] Lastly, the exchange coupling between RE-$4f$ and TM-$3d$ moments is activated via the RE-$5d$ orbitals, which enhances the MAE by entangling the RE and TM sublattices. The intra-atomic $4f$-$5d$ coupling on the RE sites is strongly ferromagnetic, while the inter-atomic $3d$-$5d$ between TM and RE atoms is AFM. This leads to the AFM coupling between the spin moments of the TM-$3d$ and RE-$4f$ moments, as observed in most $4f$-$3d$ magnets.[165] In this regard, light RE elements are more preferred[166] as permanent magnets, because the orbital moments of heavy RE elements will reduce the magnetization as they are parallel to the spin moments following the third Hund's rule.

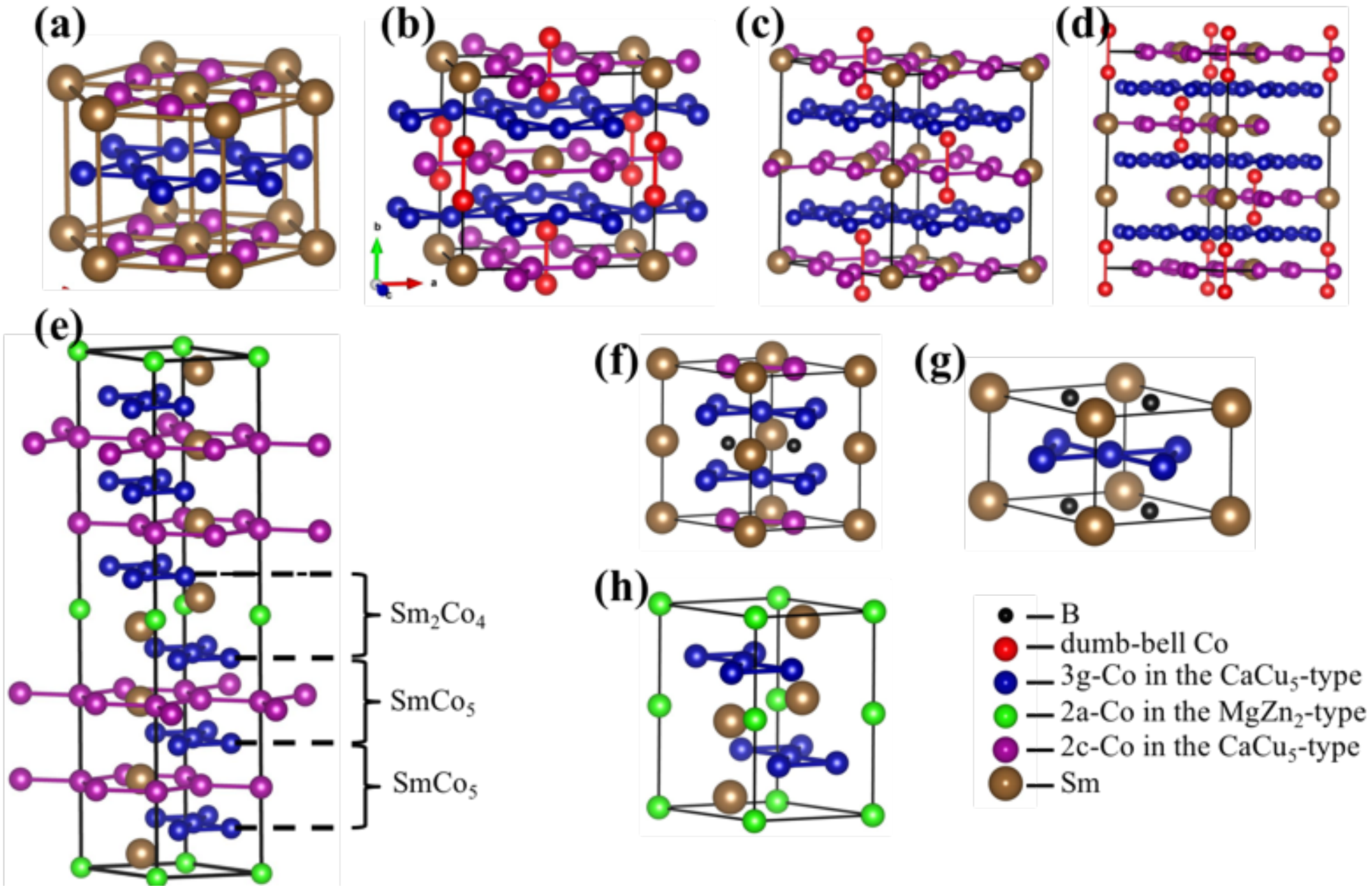

FIG. 5: Homologous structures of Sm-Co intermetallic compounds. (a) $SmCo_5$ ($CaCu_5$-type), (b) $SmCo_{12}$ ($ThMn_{12}$-type), (c) $Sm_2Co_{17}$ ($Th_2Ni_{17}$-type, 2:17H), (d) $Sm_2Co_{17}$ ($Th_2Zn_{17}$-type, 2:17R), (e) $Sm_2Co_7$ ($Ce_2Ni_7$-type), (f) $SmCo_4B$ ($CeCo_4B$-type), (g) $SmCo_3B_2$ ($Co_3GdB_2$-type), and (h) $SmCo_2$ ($MgZn_2$-type, C14 Laves). (a-d) are the members of the series $RE_{m-n}TM_{5m+2n}$ (n/m¡1/2), with (m, n) = (1, 0), (2, 1), hexagonal-(3, 1), and rhombohedra-(3, 1), respectively. (a) and (e) are the members of the series $RE_{n+2}TM_{5n+4}$ with n = ∞ and 2, respectively. (a), (f), and (g) are the members of the series $RE_{n+1}TM_{3n+5}B_{2n}$ with n = 0, 1, and ∞, respectively.

The multiple sublattice couplings have been utilized to construct atomistic spin models to understand the finite temperature magnetism of 4f-3d materials, such as $RECo_5$[167] and recently $Nd_2Fe_{14}B$.[168] Such models are ideally predictive, if the parameters are obtained from accurate electronic structure calculations based on DFT. One typical problem of such phenomenological modelling is the ill-description of the temperature dependency of the physical properties, *e.g.*, the magnitude of moments is assumed to be temperature independent.[116] This issue cannot be solved using the standard DFT which is valid at 0 K. For instance, both the longitudinal and transversal fluctuations are missing in conventional DFT calculations. Additionally, for the $4f$-$3d$ magnets, the MAE and the inter-sublattice exchange energy are of the same order of magnitude, thus the experimental measurement along the hard axis does not correspond to a simplified (anti-)parallel configuration of 4f and 3d moments.[163] This renders the traditional way of evaluating the MAE by total energy differences not applicable.[169].

Another challenging problem is how to treat the strongly correlated $4f$-electrons. Taking $SmCo_5$ as an example, there is clear experimental evidence that the ground state multiplet J=5/2 is mixed with the excited multiplets J=7/2 and J=9/2, leading to strong temperature dependence of both spin and orbital moments of Sm.[170] The problem arises how to perform DFT calculations with several multiplets, which requires more than one Slater determinants. Furthermore, Ce is an abundant element which is preferred for developing novel RE-lean permanent magnets. The $4f$ electrons in trivalent Ce can be localized or delocalized, leading to the mixed valence state, as represented by the abnormal properties of $CeCo_5$.[171]. Such features have been confirmed by the X-ray magnetic circular dichroism (XMCD) measurements in a series of Ce-based compounds.[172]

Therefore, the strongly correlated nature of $4f$-electrons calls for a theoretical framework which is beyond the standard DFT, where a consistent theoretical framework being able to treat the correlated 4f-electrons with SOC and spin fluctuations consistently is required. Most previous DFT calculations[173,174] treated the $4f$-electrons using the open-core approximation which does not allow the hybridization of $4f$-shell with the valence bands. Moreover, the spin moment of RE elements are usually considered as parallel to that of the TM atoms.[175] It is worthy mentioning that Patrick and Staunton proposed a relativistic DFT method based on the disordered local moment approximation to deal with finite temperature magnetism where the self-interaction correction (SIC) is used for the $4f$-electrons.[116] It is observed that the resulting total moment (7.13 $\mu_B$) of $SmCo_5$ is significantly underestimated compared to the experimental value of about 12 $\mu_B$.[170] In this regard, such methods such as SIC and DFT+U[76] based on a single Slater determinant may not be sufficient to capture all the features of the correlated $4f$-shells, *e.g.*, additional non-spherical correction is needed to get the the correct orbital moments.[176]

It is believed that an approach based on multiconfigurational wave functions including SOC and accurate hybridization is needed, in order to treat many close-lying electronic states for correlated open shells (*i.e.* $d$- and $f$-shells) driven by the interplay of SOC, Coulomb interaction, and crystal fields with comparative magnitude. From the quantum chemistry point of view, the full configurational interaction method is exact but can only be applied on cases with small basis sets. The complete active space self-consistent field method based on optimally chosen effective states is promising, where a multiconfigurational approach can be further developed with SOC explicitly considered.[177] Such methods have been successfully applied on the evaluation of MAE for magnetic molecules,[178] and recently extended to the four-component relativistic regime.[179] A similar method based on exact diagonalization with optimized bath functions has been developed as a potential impurity solver for DMFT calculations.[180] Thus, with the hybridization accounted for at the DFT level, such impurity solvers based on the multiconfigurational methods will make DFT+DMFT with full charge self-consistency a valuable solution to develop a thorough understanding of the magnetism in $4f$-$3d$ intermetallic compounds. Whereas previous DFT+DMFT calculations on $GdCo_5$[181] and $YCo_5$[182] have been performed using primitive analytical impurity solvers. It is noted that SOC is off-diagonal for the Anderson impurities with real spherical harmonics as the basis, leading to possible severe sign problem when the quantum Monte Carlo methods are used as the impurity solver. For such cases, either the hybridization function is diagonalized to obtain an effective basis such as demonstrated for iridates in the $5d^5$ J=1/2 states[183,184], or the variational cluster approach can be applied.[185,186] A recently developed bonding-antibonding basis offers also a possible general solution to perform DFT+DMFT on such materials with strong SOC.[187]

From the materials point of view, compounds including Fe or Co with crystal structures derived from the $CaCu_5$-type are particularly important. It is noted that $Nd_2Fe_{14}B$ has also a structure related to the $CaCu_5$-type, but its maximal permanent magnet performance has been reached to the theoretical limit, leaving not much space to improve further.[188] Interestingly, as shown in Fig. 5, various crystal structures can be derived from the $CaCu_5$-type structure. Taking Sm-Co as an example, a homologous series $RE_{m-n}TM_{5m+2n}$ can be obtained, where n out of m Sm atoms are substituted by the dumbbells of Co-Co pairs. For instance, the $ThMn_{12}$-type structure (Fig. 5b) corresponds to (m=2, n=1), *i.e.*, one Sm out of two is replaced by a Co-Co pair. For $Sm_2Co_{17}$ (m=3, n=1), both the hexagonal $Th_2Ni_{17}$-type (2:17H) (Fig. 5c) and rhombohedra $Th_2Zn_{17}$-type (2:17R) (Fig. 5d) structures can be obtained depending on the stacking of Co dumbbells. Another homologous series $RE_{n+2}TM_{5n+4}$ can be obtained by intermixing the $SmCo_2$ ($MgZn_2$-type) (Fig. 5h) and the $SmCo_5$ phases, leading to $Sm_2Co_7$ (Fig. 5e) with n=2. Lastly, by substituting B for preferentially Co on the 2c-sites, yet another homologous series $RE_{n+1}TM_{3n+5}B_{2n}$ can be obtained, leading to $SmCo_4B$ (n=1, Fig. 5f), and $SmCo_3B_2$ (n=∞, Fig. 5g). For all the derived structures, the Kagome layers formed by the Co atoms on the 3g sites of $SmCo_5$ are slightly distorted, while the changes occur mostly in the $SmCo_2$ layers due to large chemical pressure.[189]

Such a plenty of phases offer an arena to understand the structure-property relationship and to help us to design new materials, particularly the mechanical understanding based on the local atomic structures. For instance, $Sm_2Co_{17}$ has uniaxial anisotropy while all the other early RE-Co 2:17 compounds show a planar behaviour.[190] This is caused by the competition between the positive contribution from the $2c$-derived Co and the negative contribution from the dumbbell Co pairs.[190] As a matter of fact, the $2c$-Co atoms in the original $CaCu_5$-type structure have both enhanced spin and orbital moments.[191] Thus, insight on the structure-property relationship at the atomic level with defined driving forces from the electronic structure will be valuable to engineer new permanent materials.

One particularly interesting class of compounds which has drawn intensive attention recently is the Fe-rich metastable materials with the $ThMn_{12}$-type structure.[188] There have been many compounds $REFe_{11}X$ or $REFe_{10}X_2$ stabilized by substituting a third element such as Ti and V[192]. Moreover, motivated by the successful preparation of $NdFe_{12}$[193] and $SmFe_{12}$[194] thin films with outstanding permanent magnet properties, there have been many follow-up experimental and theoretical investigations[188]. The key problem though is to obtain bulk samples with large coercive field, instead of thin films fabricated under non-equilibrium conditions. In order to guide experimental exploration, detailed thermodynamic optimization of the multicomponent phase diagram is needed, where the relative formation energies in most previous DFT calculations are of minor help, *e.g.*, the prediction of $NdFe_{11}Co$[195] is disproved by detailed experiments.[196] Another challenge for the metastable Fe-rich compounds is that the $T_C$ is moderate. This can be cured by partially substituting Co for Fe. Additionally, interstitial doping with H, C, B, and N can also be helpful, *e.g.*, $Sm_2Fe_{17}N_{2.1}$ has a $T_C$ as high as 743 K, increased by 91% compared to pristine $Sm_2Fe_{17}$.[197] Interestingly, the MAE for Co-rich $ThMn_{12}$-type compounds shows an interesting behaviour. For instance, for both $RECo_{11}Ti$ and $RECo_{10}Mo_2$, almost all compounds have uniaxial anisotropy, except that Sm-based compounds show a basal-plane MAE.[198] This echoes that only $Sm_2Co_{17}$ has a uniaxial anisotropy.[190] This can be easily understood based on the crystal structure as shown in Fig. 5(b&d), where the uniaxial c-axis of the $ThMn_{12}$-type is parallel to the Kagome plane of 3g-Co atoms, but perpendicular in the case of 2:17 types. That is, from the MAE point of view, the 1:12 structure conjugates with the 2:17 structures. Thus, we suspect that substitutional and interstitial doping can be utilized extensively to engineer MAE of the 2:17 and 1:12 compounds.

Overall, although the MAE is numerically expensive to evaluate, it is a straightforward task to perform calculations with SOC considered which has been reliably implemented in many DFT codes. The key issue to perform HTP screening of permanent magnets is to properly account for the magnetism such as the spin fluctuations in RE-free and strongly correlated $4f$ electrons in RE-based materials. In addition, a systematic investigation of the intermetallic structure maps and the tunability of crystal structures via interstitial and substitutional doping will be valuable to design potential compounds. Cautions are deserved particularly when trying to evaluate the MAE for doped cases where the supercells might impose unwanted symmetry on the MAE, and also that the FM ground state should be confirmed before evaluating the MAE.

### F. Magneto-structural transitions

In general, the functionalities of ferroic materials are mostly driven by the interplay of different ferroic orders such as ferroelasticity, ferromagnetism, and ferroelectricity, which are coupled to the external mechanical, magnetic, and electrical stimuli.[200] Correspondingly, the caloric effect can be induced by the thermal response as a function of generalized displacements such as magnetization, electric polarization, strain (volume), driven by the conjugate generalized forces such as magnetic field, electric fields, and stress (pressure), leading to magnetocaloric, electrocaloric, and elastocaloric (barocaloric) effects.[201] In order to obtain optimized response, such materials are typically tuned to approach/cross a FOPT. Therefore, it is critical to understand the dynamics of FOPT in functional ferroic materials. Most FOPTs occur without long-range atomic diffusion, *i.e.*, of the martensitic type, leading to intriguing and complex kinetics. For instance, such solid-solid structural transitions are typically athermal, *i.e.*, occurring not at thermal equilibrium, due to the high energy barriers and the existence of associated metastable states. This leads to the broadening of the otherwise sharpness of FOPTs with respect to the driving parameters and hence hysteresis.[33] Therefore, mechanistic understanding of such thermal hysteresis entails accurate evaluation of the thermodynamic properties, where successful control can reduce the energy loss during the cycling.[202]

Two kinds of fascinating properties upon such FOPTs in ferroic functional materials are the shape memory[203] and caloric effects,[201] *e.g.*, MSME and MCE in the context of magnetic materials. Note that MCE can be induced in three different ways, namely, the ordering of magnetic moments (second order phase transition),[204] the magneto-structural phase transition,[205] or the rotation of magnetization,[206] all driven by applying the external magnetic fields. Taking Gd[207] with second-order phase transition as an example, under adiabatic conditions (*i.e.*, no thermal exchange between the samples and the environment) where the total entropy is conserved, applying (removing) the external magnetic fields leads to a reduction (increase) in the magnetic entropy, as the magnetic moments get more ordered (disordered), leading to increased (decreased) temperature. That is, the temperature change is mostly due to the entropy change associated with the ordering/disorder of the magnetic moments. In contrast, for materials where MCE is driven by the magneto-structural transitions of the first-order nature, additional enthalpy difference between two phases before and after the transition should be considered, although the total Gibbs free energy is continuous at the transition point. In this case, the resulting contribution to the entropy change comprise the lattice and magnetic parts, which can be cooperative or competitive. A prototype system is the Heusler Ni-Mn-Sn alloys with the inverse magnetocaloric effect.[208]

Therefore, to get enhanced magnetocaloric performance which is best achieved in the vicinity of the phase transitions, the ideal candidates should display (1) large change of the magnetization $\Delta M$, trigged by low magnetic field strengths. This makes such materials also promising to harvest waste heat based on the thermomagnetic effect,[209] (2) tunable transition temperature (*e.g.*, via composition) for the first-order magneto-structural transitions, because the transition is sharp so that a wide working temperature range can be achieved by successive compositions, and (3) good thermal conductivity to enable efficient heat exchange.

To search for potential candidates with significant MCE, it is critical to develop mechanistic understanding of the existing cases. For instance, $Ni_2MnGa$ is the only Heusler compounds with the stoichiometric composition that undergoes a martensitic transition at 200K between the $L2_1$ austenite and $DO_{22}$ martensitic phases (cf. Fig. 10), which can be optimized as magnetocaloric materials. DFT calculations show that the martensitic transition can be attributed to the band Jahn-Teller effect,[210] which is confirmed by neutron[211] and photoemission measurements.[211,212] Detailed investigation to understand the concomitant magnetic field induced strain as large as 10%[211] reveals that there exists a premartensitic phase between 200K and 260K[211] and that both the martensitic and premartensitic phases have modulated structures. It is still under intensive debate about the ground state structure and its origin. Kaufman *et al.* introduced the adaptive martensitic model, arguing that the ground state is the $L1_0$ type tetragonal phase which develops into modulated structures via nanotwinning.[213,214] On the other hand, Singh *et al.* argued based on the synchrotron x-ray measurements that the 7M-like incommensurate phase is the ground state, which is caused by the phonon softening.[215]

To shed light on the nature of the magneto-structural coupling, the Gibbs free energy $G(T, V; P)$ (Eq. (IV D)) as a function of temperature and volume $(T, V)$ for different phases $P$. According to Eq. IV D, the electronic contribution to $F^{\mathrm{ele.}}$ can be easily obtained using the methods as detailed in Ref. 216. To evaluate the vibrational term $F^{\mathrm{lat.}}$, the quasi harmonic approximation can be applied,[217] which has been implemented as a standard routine in the Phonopy code.[218] It is noted that the harmonic approximation is only valid to get the local minima on the potential energy surface, whereas proper consideration on the transition paths and kinetics requires accurate evaluation of the anharmonic effects.[219] However, when the spin-phonon interaction becomes significant as in the paramagnetic states, it is still a challenging task to evaluate the phonon spectra and hence the lattice free energies. The spin space average technique has been developed and applied for bcc Fe[119] and we believe the recent implementation in the DMFT regime is very promising.[125] For the magnetic free energy $F^{\mathrm{mag.}}$, the classical Monte Carlo simulations, as usually done based on the Heisenberg model (Eq. (IV B)) with DFT derived exchange coupling $J_{ij}$ to obtain the Curie temperature, are not enough, as the specific heat at 0K stays finite instead of the expected zero, due to the continuous symmetry of spin rotations. Additionally, quantum Monte Carlo suffers from the sign problem as $J_{ij}$ is long range and changes its sign with respect to the distance. One solution is to rescale the classical Monte Carlo results, as done for bcc Fe.[105] An alternative is to map the free energy based on the fitted temperature dependence of the magnetization.[220] This has been applied successfully to get the martensitic transition temperature and the stability of premartensitic phases in Ni-Mn-Ga system.[221] The recently developed spin dynamics with quantum thermostat[106], the Wang-Landau technique to sample the thermodynamic density of states in the phase space directly[222], and the atomistic spin-lattice dynamics are interesting to explore in the future.[223] Last but not least, magnetic materials with noncollinear magnetic ordering and thus vanishing net magnetization can also be used for caloric effects such as barocaloric[224] and elastocaloric[225] effects, it is essential to evaluate the free energy properly which has been investigated recently.[112]

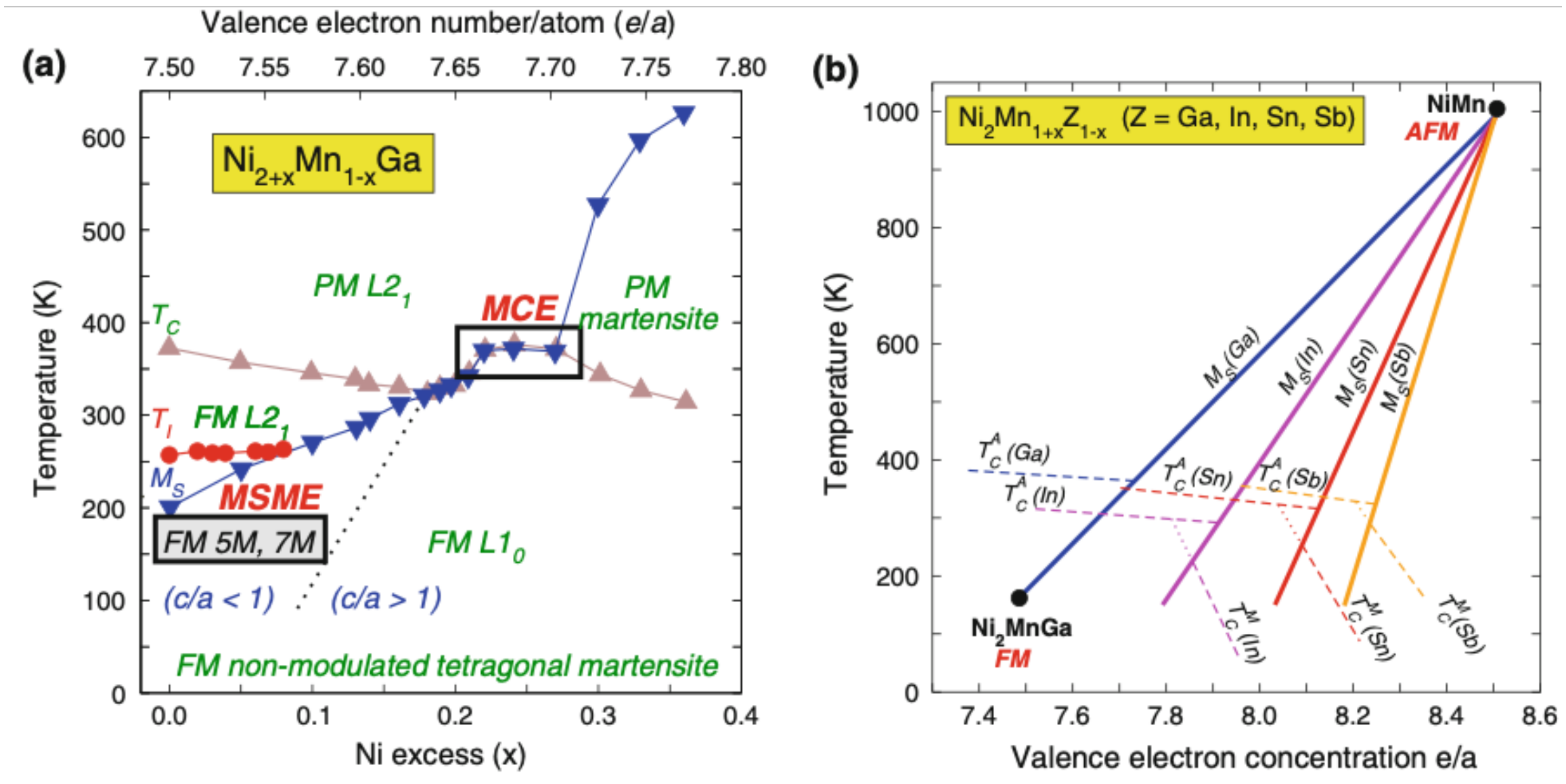

FIG. 6: Phase diagram of Ni-Mn-X.[199] Copyright requested.

The magnetocaloric performance of stoichiometric $Ni_2MnGa$ is not significant, as the martensitic transition occurs between the FM $L2_1$ and modulated martensitic phase. As shown in Fig. 6(a), excess Ni will bring the martensitic phase transition temperature and $T_C$ together, and hence enhance the magnetocaloric performance.[226] Interestingly, such a phase diagram can be applied to interpolate the transitions in several Ni-Mn-X (X = Ga, In, Sn, Sb) alloys, where the dependence with respect to the valence electron concentration can be formulated (Fig. 6(b)). The valence electron concentration accounts the weighted number of the $s/p/d$ electrons, which can be tuned by excess Mn atoms. For instance, the martensitic transition temperature for the Ni-Mn-X alloys interpolates between the stoichiometric $Ni_2MnX$ and the AFM NiMn alloys. In such cases, the martensitic transition occurs usually between the FM $L2_1$ and the paramagnetic $L1_0$ phases, leading to the metamagnetic transitions and hence giant magnetocaloric performances.[205,208] The magnetocaloric performance can be further enhanced by Co doping,[227–229] which enhances the magnetization of the austenite phase. However, Co-doping modifies the martensitic transition temperature significantly, leading to a parasitic dilemma.[230]

From the materials point of view, there are many other classes of materials displaying significant magnetocaloric performance beyond the well-known Heusler compounds, as summarized in Fig. 7.[231] Three of them, *e.g.*, Gd alloys,[232] La-Fe-Si,[233,234] and MnFe(Si,P)[235] have been integrated into devices, where $La(FeSi)_{13}$ is considered as the most promising compounds for applications, driven by the first-order metamagnetic phase transition at $T_C$.[236] For such materials, research has been focusing on the fine tuning of the magnetocaloric performance by chemical doping. As discussed above, the key problem to overcome when optimizing and designing MCE materials is to reduce the hysteresis in order to minimize the energy losses,[237] this is a challenging task for HTP design where the Gibbs free energies for both the end phases and the intermediate phases are needed in order to have a *complete* thermodynamic description of the phase transition. In this regard, a particularly promising idea is to make use of the hysteresis rather than to avoid it, which can be achieved in the multicaloric regime engaging multiple stimuli. This has been demonstrated recently in Ni-Mn-In system[238] combining magnetic fields and uniaxial strains.

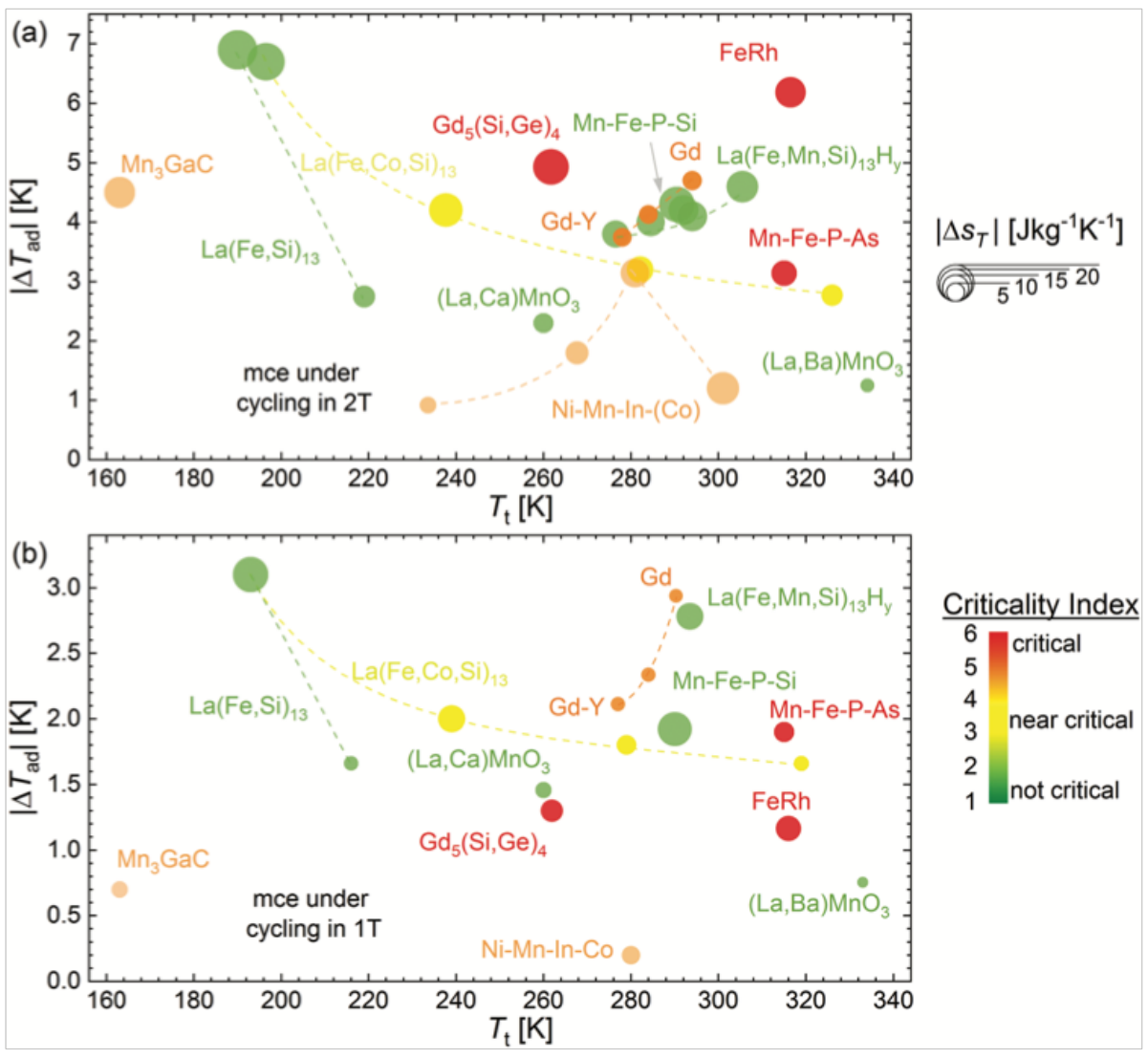

FIG. 7: Library of MCE materials[231] Copyright requested.

Therefore, designing novel MCE materials entails joint theoretical and experimental endeavour. HTP calculations can be applied to firstly search for compounds with polymorphs linked by group-subgroup relationships, and then to evaluate the Gibbs free energy for the most promising candidates. Nevertheless, for such materials with first-order magneto-structural transitions, hysteresis typically arises as a consequence of nucleation, in caloric materials it occurs primarily due to the domain-wall pinning, which is the net result of long-range elastic strain associated with phase transitions Thus, experimental optimization on the microstructures[239] (Sect. VI A) is unavoidable in order to obtain a practical material ready for technical applications.

## G. Spintronics

Till now we have been focusing on the physical properties of the equilibrium states, whereas the non-equilibrium transport properties of magnetic materials leads to many interesting properties, where spintronics is one class of the most interesting phenomena. In contrast to the conventional electronic devices, the spin degree of freedom of electrons has been explored to engineer more energy efficient devices.[35] The first generation of spintronic applications are mostly based on the FM materials, initiated by the discovery of giant magnetoresistance (GMR) in magnetic multilayers in 1988,[240,241] *e.g.*, with the MR ratio as large as 85% between parallel and antiparallel configurations of Fe layers separated by the Cr layers in-between. The underlying mechanism of GMR can be understood based on the two-current model,[242] where spin-polarized currents play a deterministic role. Two important follow-up development of GMR are the tunnelling magnetoresistance (TMR)[243] and spin transfer torque (STT).[244,245] The TMR effect is achieved in multilayers where the ferromagnetic metals are separated by large gap insulators such as MgO, and the MR ratio can be as large as 600% in CoFeB/MgO/CoFeB by improving the surface atomic morphology.[246] In the case of STT, by driving electric currents through the reference layer, finite torque can be exerted on the freestanding layers and hence switch their magnetization directions. This enables also engineering spin transfer oscillator,[247] racetrack memory,[248] and nonvolatile RAM.[249]

The second generation of spintronics comprise phenomena driven by SOC, dubbed spin-orbitronics, which came around 2000.[36] For ferromagnetic materials, SOC gives rise to the anisotropic magnetoresistance (AMR)[250], where the resistivity depends on the magnetization direction, in analog to GMR but without the necessity to form multilayers. The most essential advantage of spin-orbitronics is to work with spin current (*i.e.*, a flow of spin angular momentum ideally without concomitant charge current), instead of the spin polarized current.[251] Correspondingly, the central subjects of spin-orbitronics are the generation, manipulation, and detection of spin current. Note that after considering SOC, spin is not any more a good quantum number, thus it is still an open question how to define the spin current properly.[252]

Two most important phenomena for spin-orbitronics are the spin Hall effect (SHE)[291] and spin-orbit torque (SOT).[292] SHE deals with the spin-charge conversion, where transversal spin current can be generated by a longitudinal charge current. It has been observed in paramagnetic metals such as Ta[293] and Pt,[294] FM metals like FePt,[295] AFM metals including MnPt,[296] $Mn_{80}Ir_{20}$,[297] and $Mn_3Sn$,[298] semiconductors such as GaAs,[299] and topological materials such as $Bi_2Se_3$[300] and TaAs.[301] It is noted that the reciprocal effect, *i.e.*, the inverse SHE (iSHE), can be applied to detect the spin current by measuring the resulting charge current.

TABLE II: Selected systems implementing spin-orbitronics and AFM spintronics. Copyright requested for the figures.

| | spin transport | spin-orbit torque | spin pumping |
|---|---|---|---|
| | 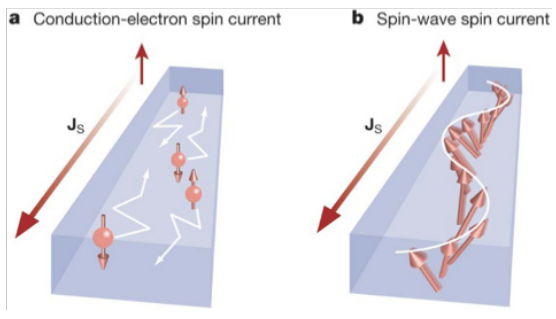<br>spin current[253] | 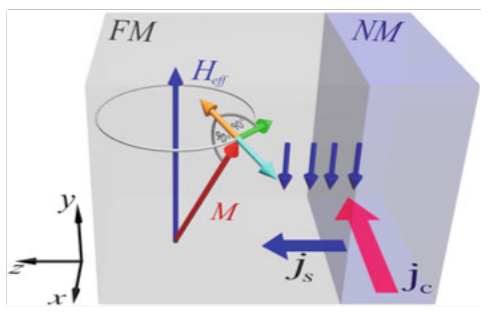<br>SOT in NM\|FM bilayers[254] | 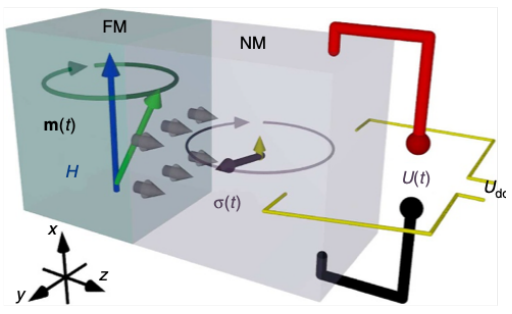<br>spin pumping in FM\|NM bilayers[255] |
| FM metal | review[256] | Pt\|Co[257,258]<br>Ta\|CoFeB[258,261]<br>$SrIrO_3$\|NiFe[263] (oxides)<br>$WTe_2$\|NiFe[265] (out-of-plane)<br>NiMnSb[266] (bulk) | NiFe\|Pt[255,259,260]<br>NiFe\|GaAs[262] (metal\|semiconductor)<br>NiFe\|YIG[264] (metal\|insulator) |
| FM insulator | YIG[267] | GaMnAs\|Fe[268]<br>GaMnAs[270] (bulk) | GaMnAs\|GaAs[269]<br>YIG\|Pt[253,271] |
| AFM metal | | IrMn\|CoFeB[272]<br>IrMn\|NiFe[274]<br>MnPt\|Co/Ni[275,276]<br>$Mn_3Ir$\|NiFe[277]<br>CuMnAs[278] (bulk)<br>$Mn_2Au$[279,280] (bulk) | NiFe\|IrMn[273] |
| AFM insulator | $Cr_2O_3$[281]<br>$Fe_2O_3$[287] | Pt\|NiO[282–284]<br>Pt\|TmIG[288]<br>$Bi_2Se_3$\|NiO\|NiFe[290] | YIG\|NiO[285,286]<br>YIG\|CoO[289] |

SOT reflects with the interaction of magnetization dynamics and the spin current, leading to magnetization switching and spin-orbit pumping. Following the Landau-Lifshitz-Gilbert (LLG) equation,[292]

$$\frac{d\mathbf{e_m}}{dt} = \underbrace{-\gamma \mathbf{e_m} \times \mathbf{B}}_{\text{precession}} + \underbrace{\alpha \mathbf{e_m} \times \frac{d\mathbf{e_m}}{dt}}_{\text{relaxation}} + \underbrace{\frac{\gamma}{M_s}\mathbf{T}}_{\text{torque}}, \tag{14}$$

where $\gamma$, $\alpha$, and $M_s$ mark the gyromagnetic ratio, the Gilbert damping parameter, and the saturation magnetization, respectively. $\mathbf{e_m}$, $\mathbf{B}$, and $\mathbf{T}$ indicate the magnetization unit vector, the effective field, and the total torque, respectively. The torque $\mathbf{T}$ is perpendicular to $\mathbf{m}$, which can be generally expressed as[292]

$$\mathbf{T} = \underbrace{\tau_{\text{FL}} \mathbf{e_m} \times \boldsymbol{\epsilon}}_{\text{field-like}} + \underbrace{\tau_{\text{DL}} \mathbf{e_m} \times (\mathbf{e_m} \times \boldsymbol{\epsilon})}_{\text{damping-like}}, \tag{15}$$

where $\boldsymbol{\epsilon}$ is the unit vector of the torque. Note that the *field-like* and *damping-like* terms act on the magnetization like the precession and relaxation terms in the LLG equation (Eq. (IV G)).

To induce finite SOT, effective non-equilibrium spin polarization is required, which can be obtained via (a) SHE and (b) Edelstein effect[302] (*i.e.*, inverse spin Galvanic effect(iSGE)[303]). From symmetry point of view, the occurrence of SOT requires noncentrosymmetric symmetry, thus the heterostructures of NM/FM materials provide a rich playground to investigate SOT, where paramagnetic materials with significant SHE have been the dominant source for inducing spin injection and SOT in FM materials. For instance, SOT has been investigated in Ta|CoFeB,[261] Pt|Co,[257,258], $SrIrO_3 mid$NiFe,[263] and $Bi_2Se_3|BaFe_{12}O_{19}$.[304] On the other hand, the Edelstein effect requires Rashba splitting, which can also be applied for spin-charge conversion and hence SOT. For instance, the spin-charge conversion has been demonstrated for 2DEG at the interfaces of $LaAlO_3|SrTiO_3$,[305] Ag|Bi,[306], and $Cu|Bi_2O_3|NiFe$.[307] Correspondingly, SOT is recently observed in $NiFe|CuO_x$ driven by the Rashba splitting at the interfaces.[308] We note that the SHE-SOT and iSGE-SOT (Edelstein) are entangled, *e.g.*, can be competitive or cooperative with each other, as demonstrated in (GaMn)As|Fe.[268]

It is noteworthy pointing out that from the symmetry perspective,[258] the SOT generated at the interfaces of FM metals and heavy metals lies in-plane for both the anti-damping and field-like contributions, thus can only switch the in-plane magnetization direction efficiently. Nevertheless, it is demonstrated that in $WTe_2$|NiFe,[265] out-of-plane SOT can be induced by current along a low-symmetry axis, which is promising for future SOT switching of FM materials with perpendicular magnetic anisotropy. Although the SOT discussed above takes place at the interfaces, it does not mean that SOT is prohibited in bulk materials. For instance, SOT has been confirmed in half Heusler NiMnSb.[266] Thus, the key is to break the inversion symmetry in order to obtain finite SOT.

As SOT can be summarized as torque induced by spin current, the Onsager reciprocal effect is spin pumping, *i.e.*, the generation of spin current via induced magneti-zation dynamics. Usually the spin pumping is induced by generating non-equilibrium magnetization dynamics using the ferromagnetic resonance (FMR) and hence creates a pure spin current which can be injected into the adjacent NM layers (*i.e.*, spin sink) without a charge flow under zero bias voltage. For instance, it is demonstrated that spin pumping can be achieved in NiFe|Pt bilayers,[259] with Pt being the spin sink, which can be conveniently detected via the iSHE. In this regard, many materials have been considered as spin sinks, such as heavy metals like Au and Mo[260] and semiconductors as GaAs.[262] An intriguing aspect of spin pumping is that the polarization of the spin current is time-dependent, leading to both *dc*- and *ac*-components.[309] It is demonstrated that the ac-component is at least one order of magnitude larger than the *dc*-component for NiFe|Pt,[255] which can lead to future *ac* spintronic devices.

An emergent field of great interest is AFM spintronics, which possess several advantages over the FM counterpart.[37,310,311] For instance, there is no stray field for AFM materials thus the materials are insensitive to the neighboring unit, which allows denser integration of memory bits. Moreover, the typical resonance frequency of FM materials is in the GHz range which is mostly driven by MAE,[312] while that for AFM materials can be in the THz range due to the exchange interaction between the moments.[313] This leads to ultrafast magnetization dynamics in AFM materials. As a matter of fact, almost all the spintronic phenomena existing in FM materials have been observed in AFM materials, such as AMR in FeRh[314] and MnTe,[315] SHE in MnPt,[296] and tunnelling AMR.[316,317] Nevertheless, one major challenge for AFM spintronics is how to control and to detect the AFM ordering. It turns out that SOT can be applied to switch the magnetic ordering of AFM compounds. For instance, the iSGE-SOT can lead to staggered Néel spin-orbit torques, creating an effective field of opposite sign on each magnetic sublattice. The SOT induced switching has been observed in CuMnAs[278] and $Mn_2Au$.[279,280] It is noted that the staggered SOT has special requirement on the symmetry of the materials, *i.e.*, the inversion symmetry connecting the AFM magnetic sublattices in the crystal structure is broken by the magnetic configuration.[37] Thus, it is an interesting question whether there are more materials with the demanded magnetic configurations for bulk AFM spintronics. To detect the AFM magnetic ordering, spin Hall magnetoresistance effect can be used.[318]

One particularly interesting subject is the SOT and spin pumping for magnetic junctions combining FM and AFM materials. For instance, as AFM materials display also significant SHE, SOT has been observed in, *e.g.*, IrMn|CoFeB,[272] IrMn|NiFe,[274] MnPt|Co,[275] and $Mn_3Ir$|NiFe.[277] It is demonstrated in MnPt|Co/Ni bilayers, the resulting SOT switching behaves like an artificial synapses, and a programmable network of 36 SOT devices can be trained to identify 3×3 patterns.[276] This leads to an intriguing application of spintronic devices for neuromorphic computation.[319] Furthermore, as spin pumping depends on the magnetic susceptibility at the interfaces and thus the magnetic fluctuations, it is observed that the spin pumping in NiFe|IrMn bilayers is significantly enhanced around the AFM phase transition temperature of IrMn,[273] consistent with the theoretical model based on enhanced interfacial spin mixing conductance.[320] In addition to enhance the spin pumping efficiency, such types of experiments can be applied to detect the AFM ordering without being engaged with the neutron scattering facilities.

Another emerging field of spintronics is magnonic spintronics[321] realized based on both FM and AFM insulators, as each electron carries an angular momentum of $\hbar/2$ whereas each magnon as the quasiparticle of spin wave excitations carries an angular momentum of $\hbar$. Theoretically, it is predicted that at the interfaces of NM|FM-insulator bilayers, spin gets accumulation which is accompanied by the conversion of spin current to magnon current.[322] In the case of FM (ferrimagnetic) insulators, YIG|Pt is a prototype bilayer system to demonstrate the transmission of spin current at the interfaces[253] and spin pumping.[271,323] It is also demonstrated the SOT can be applied to switch the magnetization directions of FM insulators, as in Pt|TmIG.[288] The SOT and spin pumping have also been observed in AFM insulators. For instance, SOT leads to the switching of AFM ordering of NiO in Pt|NiO bilayers.[282–284] The origin of switching can be attributed to the non-staggered SHE-SOT which acts as anti-damping-like torque exerted by a spin accumulation at the interface, but this is a question under intensive debate.[284] Furthermore, spin pumping can be induced in bilayers of FM|AFM-insulators such as YIG|NiO[285]. One intriguing aspect is the non-locality of magnetic spin current which can propagate 100 nm in NiO.[286] In general, magnon number is not conserved, leading to finite transport length scale for magnon mediated spin current determined by the Gilbert damping coefficient which is an intrinsic character of compounds. In contrast to the FM materials,[256] AFM materials allow long range spin transport, as observed in $Cr_2O_3$[281] and $Fe_2O_3$[324] which might be attributed to the spin superfluidity.[325] Particularly, the damping of magnons can be compensated by SOT, which leads to zero damping as observed very recently in YIG.[267] Lastly, it is a fascinating idea to combine SOT and long diffusion length of spin current in AFM materials, *e.g.*, switching magnetism using the SOT carried by magnons in $Bi_2Se_3$|NiO|NiFe.[290]

In short, spintronics is a vast field but there is an obvious trend that more phenomena driven by SOC are being investigated. Moreover, materials to generate, manipulate/conduct, and detect spin current should be integrated for devices, thus the interfacial engineering is a critical issue. Following Table. IV G, both FM and AFM metals/insulators can be incorporated into the spintronic devices, resulting in flexibility but also time consuming combinatorial optimizations. For instance, a few FM metals like Fe/Co/Ni, permalloy (NiFe), and CoFeB have been widely applied in the spintronic heterostructures, as the experimental techniques to fabricate such systems are well established. In this regard, there is a significant barrier between theoretical predictions and experimental implementations, though HTP calculations can be carried out in a straightforward way (cf. Sect. V).

### H. Magnetic topological materials

Pioneered by the discovery of integer[326,327] and fractional[328,329] quantum Hall effects, a resurgence of research on materials with nontrivial topological nature has began, as exemplified by the 2D topological insulators (TIs) predicted in graphene[330] and HgTe quantum wells,[331] and shortly afterwards confirmed experimentally in many 2D and 3D systems.[332,333] The key fundamental concept lies on the Berry phase, which measures the global geometric phase of the electronic wave functions accumulated through adiabatic evolutions.[334] Such materials host many fascinating properties such as quantum spin Hall effect (QSHE), which are promising for future applications such as topological spintronics.[332,333] Particularly, introducing the magnetic degree of freedom not only enriches their functionalities, *e.g.*, quantum anomalous Hall effect (QAHE),[335] axion electromagnetic dynamics,[336,337] and chiral Majorana fermions,[338] but also enables more flexible tunability. The relevant materials as compiled in Tab. III, categorized based on the dimensionality of the resulting band touchings, namely, topological insulator, nodal point semimetals (including Weyl, Dirac, and manyfold nodal points[339]), and nodal line semimetals.[340]

As symmetry plays an essential role in the topological nature of such materials,[341] we put forward a brief discussion before getting into particular material systems. It is clear that the time-reversal symmetry $\Theta$ is broken for all magnetically ordered compounds, and hence they are good candidates to search for the QAHE and Weyl semimetal phases, which are topologically protected rather than symmetry protected. In addition, the Weyl node is a general concept which is closely related to the accidental degeneracy of electronic bands,[342] *e.g.*, there are many Weyl points in bcc Fe.[343] Thus, the ultimate goal is to search for compounds where the low energy electronic structure around the Fermi energy are dominated solely (or mostly) by the linear dispersions around the Weyl points.

The occurrence of other topological phases such as TI and Dirac fermions in magnetic materials requires extra symmetry. For instance, the Kramers degeneracy is required to define the $\mathcal{Z}_2$ index for nonmagnetic TI.[330] However, the electronic bands in magnetic materials are generally singly degenerate except at the time-reversal invariant momenta (TRIMs), therefore additional symmetry is required in order to restore the Kramers degeneracy. This leads to the prediction of AFM topological insulators,[344] which are protected by an anti-unitary product symmetry $\mathcal{S} = \mathcal{P}\Theta$, where $\mathcal{P}$ can be a point group symmetry[345] or a nonsymmorphic symmetry operator which reverses the magnetization direction of all moments. Furthermore, it is noted that the inversion symmetry $\mathcal{I}$ respects the magnetization direction, thus parity as the resulting eigenvalues of $\mathcal{I}$ can still be applied to characterize the topological phase if the compound is centrosymmetric. Nevertheless, the generalized $\mathcal{Z}_4$ character should be used,[346] *e.g.,* $4n{+}2$ number of odd parities from the occupied states at TRIMs corresponds to a nontrivial axion insulator. In the same way, for topological semimetallic phases,[339] additional crystalline line symmetry should be present, as will be discussed for particular materials below.

Turning now to the first class of materials exhibiting QAHE, which was predicted by Haldane.[385] Such a phenomena can be realized in magnetic topological insulators,[386] with magnetism introduced by doping as observed in $Cr_{0.15}(Bi_{0.1}Sb_{0.9})_{1.85}Te_3$ at 30 mK.[347] There have been many other proposals based on DFT calculations to predict finite temperature QAHE,[335] awaiting further experimental validations. Particularly, compounds with honeycomb lattices are a promising playground for QAHE, including functionalized graphene,[387,388] $Mn_2C_{18}H_{12}$,[389] etc. We want to point out that QAHE is defined for 2D systems,[385] an interesting question is whether there exists 3D QAHI. By making an analogue to the 3D integer quantum Hall effect,[390] the 3D QAHI can be obtained by stacking 2D QAHI on top of each other, or by stacking Weyl semimetals.[391] In the former case, if the Chern number for each $k_z$-plane (in the stacking direction) changes, one will end up with Weyl semimetals.[392] It is proposed recently that for $Ba_2Cr_7O_{14}$ and $h$-$Fe_3O_4$,[348] 3D QAHI can be obtained due to the presence of inversion symmetry.

The 2D nature of QAHE observed in Cr-doped TIs arises two questions: (1) whether the gap opening in the Dirac surface states required for QAHE can be confirmed experimentally and (2) whether it is possible to engineer axion insulators by forming heterostructures of magnetic-TI/normal-TI/magnetic-TI. For the former, a band gap of 21 meV was observed in Mn-doped $Bi_2Se_3$[393] but proved not originated from magnetism.[394] Only recently, a comparative study of Mn-doped $Bi_2Te_3$ and $Bi_2Se_3$ reveals that the perpendicular MAE for the Mn moments in $Bi_2Te_3$ is critical to induce a finite band gap of 90 meV opening in the Dirac surface states.[395] In the latter case, if the magnetization directions for two magnetic-TIs are antiparrallel and the sandwiching normal-TI is thick enough, the axion insulator phase can be achieved with quantized *ac*-response such as magneto-optical effect and topological magnetoelectric effect;[396] Whereas QAHE with quantized *dc*-response corresponds to the parallel magnetization of two magnetic-TIs.[397] Several experiments have been done in this direction with convincing results.[398–400]

Focusing from now on only the *intrinsic* (bulk, non-doped, non-heterostructure) magnetic topological mate-rials, an inspiring system is $MnBi_2Te_4$, which is confirmed to be an AFM TI recently.[349] It is an ordered phase with one Mn layer every septuple layer in the $Bi_2Te_3$ geometry and the Mn atoms coupled ferromagnetically (antiferromagnetically) within (between) the Mn layers, leading to a Néel temperature of 24 K. Thus, the nontrivial topological phase is protected by the combined symmetry $\mathcal{S} = \Theta\mathcal{T}$ with $\mathcal{T}$ being the translational operator connecting two AFM Mn-layers.[349] It is worthy mentioning that the $MnBi_2Te_4$ phase is actually the reason why the observed band gap is as large as 90 meV in Mn-doped $Bi_2Te_3$.[395] In this regard, it is still an open issue about whether the Dirac surface states of $MnBi_2Te_4$ are gapped[401–403] or gapless[354,404,405] and the corresponding magnetic origin as the gap survives above the Néel temperature.[403] Nevertheless, $MnBi_2Te_4$ stands for a family of compounds with a general chemical formula $MnBi_{2n}(Se/Te)_{3n+1}$, which are predicted to host many interesting topological phases.[406]

TABLE III: The incomplete list of magnetic topological materials. SdH: Shubnikov-de Haas

| compounds | phase | space group | measurements |
|---|---|---|---|
| $Cr_{0.15}(Bi_{0.1}Sb_{0.9})_{1.85}Te_3$ | QAHI | R$\bar{3}$m | transport at 30 mK[347] |
| $Ba_2Cr_7O_{14}$ | QAHI | R$\bar{3}$m | DFT[348] |
| $MnBi_2Te_4$ | AFM TI | R$\bar{3}$m | DFT+ARPES[349] |
| $EuSn_2P_2$ | AFM TI | R$\bar{3}$m | ARPES[350] |
| $EuIn_2As_2$ | AFM axion | $P6_3$/mmc | DFT,[351] ARPES[352,353] |
| $EuSn_2As_2$ | AFM TI | R$\bar{3}$m | DFT+ARPES[354] |
| FeSe monolayers | 2D AFM TI | P4/nmm (bulk) | DFT+ARPES+STS[355] |
| $HgCr_2Se_4$ | FM Weyl | Fd$\bar{3}$m | DFT[356,357] |
| $Co_2MnGa$ | FM Weyl/nodal-line | Fm$\bar{3}$m | DFT,[358] ARPES[359] |
| $EuCd_2As_2$ | ideal Weyl | P$\bar{3}$m1 | DFT+ARPES[360]+SdH[361] |
| GdPtBi | AFM Weyl | F$\bar{4}$3m | DFT+transport[362] |
| YbPtBi | AFM Weyl | F$\bar{4}$3m | DFT+ARPES+transport[363] |
| $Co_3Sn_2S_2$ | FM Weyl | R$\bar{3}$m | DFT+transport[364] |
| $Mn_3Sn$ | AFM Weyl | $P6_3$/mmc | DFT+APRES[365] |
| $Fe_3Sn_2$ | FM Weyl | R$\bar{3}$m | ARPES,[366] STS+QPI[367] |
| $(Y/RE)_2Ir_2O_7$ | AFM Weyl/axion TI | Fd$\bar{3}$m | DFT[368] |
| CeAlGe | Weyl | $I4_1$md | transport[369] |
| CuMnAs | AFM massive Dirac | Pnma | DFT,[370,371] transport[372] |
| GdSbTe | AFM Dirac | P4/nmm | DFT+ARPES[373] |
| $BaFe_2As_2$ | AFM massless Dirac | I4/mmm | DFT+Infrared,[374] ARPES[375] |
| $CaIrO_3$ | AFM Dirac | Pnma | transport+SdH[376] |
| FeSn | AFM massless Dirac | P6/mmm | DFT+ARPES[377,378] |
| $CaMnBi_2$ | AFM massive Dirac | P4/nmm | ARPES[379] |
| $(Sr/Ba)MnBi_2$ | AFM massive Dirac | I4/mmm | DFT+ARPES[379,380] |
| $EuMnBi_2$ | AFM massive Dirac | I4/mmm | transport+SdH[381] |
| $YbMnBi_2$ | AFM massive Dirac | P4/nmm | DFT+ARPES[382] |
| $GdAg_2$ | 2D nodal-line | – | DFT+ARPES[383] |
| GdPtTe | AFM nodal-line | P4/nmm | DFT+ARPES[373] |
| $MnPd_2$ | AFM nodal-line | Pnma | DFT[384] |

The topological materials can be designed based on the chemistry in systems with comparable crystal structures.[407,408] Taking three Eu-based compounds listed in Tab. III as an example, the electronic states around the Fermi energy are mainly derived from the In-5$s$/Sn-5$p$ and P-3$p$/As-4$p$ states, whereas the Eu-4$f$ electrons are located more than 1 eV below the Fermi energy.[350,351,354] Certainly the hybridization between the 4f and valence electronic states around the Fermi energy is mandatory in order to drive the systems into the topological phases by the corresponding magnetic ordering. Nevertheless, it is suspected that given appropriate crystal structures, there is still free space to choose a the chemical composition, *i.e.*, via substituting chemically similar magnetic RE elements to fine tune the electronic structure and thus design novel materials. This philosophy is best manifested in the class of Dirac semimetals $AMnBi_2$ (A = Ca, Sr, Bi, Eu, and Yb) as listed in Tab. III. For such compounds, the Dirac cone can be attributed to the square lattice of Bi atoms, which is common to all the compounds following the DFT prediction.[409] It is noted that the Dirac cones in all the $AMnBi_2$ compounds are massive due to SOC, but still leads to high mobility as the resulting effective mass of electrons is small.[410] Substituting Sb for Bi leads to another class of Dirac semimetals $AMnSb_2$ as predicted by DFT[411] and confirmed experimentally.[410,412–415]

As discussed above, the occurrence of Dirac points (with four-fold degeneracy) in magnetic materials demands symmetry protection. Taking CuMnAs as an example, it is demonstrated that the Dirac points is protected via the combined symmetry of $\mathcal{S} = \mathcal{I}\Theta$ as a product of inversion $\mathcal{I}$ and $\Theta$.[370] Considering additionally

SOC leads to lifted degeneracies depending on the magnetization directions because of the screw rotation operator $S_{2z}$. The same arguments applies to FeSn[377] and $MnPd_2$.[384] Several comments are in order. Firstly, it is exactly due to the combined symmetry $\mathcal{S} = \mathcal{I}\Theta$ why there exists staggered SOT for CuMnAs, leading to a descriptor to design AFM spintronic materials.[371] Secondly, the magnetization direction dependent electronic structure is a general feature for *all* magnetic topological materials, such as surface states of AFM TIs like $MnBi_2Te_4$[349] and $EuSn_2As_2$,[354] and also the nodal line and Weyl semimetals.[416] This offers another way to tailor the application of the magnetic topological materials for spintronics.[417] Thirdly, the Dirac (Weyl) features prevail in AFM (FM) materials as listed in Tab. III, thus we suspect that there are good chances to design AFM materials with high mobility of topological origin.

The relevant orbitals responsible for the topological properties for most of the materials discussed so far are $p$-orbitals of the covalent elements, where the TM-$d$ and RE-$f$ states play an auxiliary role to introduce magnetic ordering. A particularly interesting subject is to investigate the topological features in bands derived from more correlated $d/f$-orbitals, which has recently been explored extensively in ferromagnetic Kagome metallic compounds in the context of Weyl semimetals. We note that the Kagome lattice with AFM nearest neighbor coupling has been an intensively studied field for the quantum spin liquid.[418] On the one hand, Weyl semimetals host anomalous Hall conductivity[419] and negative magnetoresistance driven by the chiral anomaly.[420,421] On the other hand, due to the destructive interference of the Bloch wave functions in the Kagome lattice, flat bands and Dirac bands can be formed with nontrivial Chern numbers.[422,423] Therefore, intriguing physics is expected no matter whether the Fermi energy is located in the vicinity of the flat band or the Dirac gap, as in the $T_mX_n$ compounds (T = Mn, Fe, and Co; X = Sn and Ge).[378] Particularly, for $Fe_3Sn_2$, the interplay of gauge flux driven by spin chirality and orbital flux due to SOC leads to giant nematic energy shift depending on the magnetization direction and further the spin-orbit entangled correlated electronic structure.[367] This offers a fascinating arena to investigate the emergent phenomena of topology and correlations in such quantum materials, which can also be generalized to superconducting materials such as Fe-based superconductors.[355,374,375]

From the materials point of view, there are many promising candidates to explore. For instance, the recently discovered FM $Co_3Sn_2S_2$ shows giant anomalous Hall conductivity[364] and significant anomalous Nernst conductivity,[424], driven by the Weyl nodes around the Fermi energy. In addition, various Heusler compounds displays also Weyl features in the electronic structure, including $Co_2MnGa$,[358,359] $Co_2XSn$,[425,426] $Fe_2MnX$,[426] and $Cr_2CoAl$.[427] The most interesting class of compounds are those with noncollinear magnetic configurations. It is confirmed both theoretically and experimentally that $Mn_3Sn$ is a Weyl semimetal,[365,428] leading to significant AHC[429] and SHC.[298] It is expected that the Weyl points can also be found in the other noncollinear systems such as $Mn_3Pt$[430] and antiperovskite,[46] awaiting further investigation.

Iridates are another class of materials which are predicted to host different kind of topological phases.[431,432] Most insulating iridates contain $Ir^{4+}$ ions with five $5d$ electrons, forming $IrO_6$ octahedra with various connectivities. Due to strong atomic SOC, the $t_{2g}$ shell is split into low-lying J = 3/2 and high-lying half-filled J = 1/2 states. Therefore, moderate correlations can lead to insulating states, and such unique SOC-assisted insulating phase have been observed in Ruddlesden-Popper iridates.[433] Pyrochlore iridates $RE_2Ir_2O_7$ can also host such insulating states, which were predicted to be topologically nontrivial.[368] Many theoretical studies based on models have been carried out, supporting the nontrivial topological nature,[434435] but no smoking-gun evidence, *e.g.*, surface states with spin-momentum locking using ARPES, has been observed experimentally. Our DFT+DMFT calculations with proper evaluation of the topological character suggest that the insulating $RE_2Ir_2O_7$ are likely topologically trivial,[184] consistent with the ARPES measurements.[436] Nevertheless, proximity to the Weyl semimetal phase is suggested, existing in a small phase space upon second-order phase transition by recent experiments,[437] where magnetic fields can applied to manipulate the magnetic ordering of the RE-sublattices and thus the topological character of the electronic states.[438439440] Last but not least, it is observed that there exist robust Dirac points in $CaIrO_3$ with the post-perovskite structure,[376] leading to high mobility which is rare for correlated oxides.

Obviously, almost all known topological phases can be achieved in magnetic materials, which offer more degrees of freedom compared to the nonmagnetic cases to tailor the topological properties. Symmetry plays a critical role, *e.g.*, semimetals with nodal line[383373384] and many-fold nodal points,[339] can also be engineered beyond the Dirac nodes and AFM TIs discussed above. Unlike nonmagnetic materials where the possible topological phases depend only on the crystal structure and there are well compiled databases such as ICSD, there is unfortunately no complete database with the magnetic structures collected. Also, it is fair to say that no ideal Weyl semimetals have been discovered till now, *e.g.*, there are other bands around the Fermi energy for all the Weyl semimetal materials listed in Tab. III. Therefore, there is a strong impetus to carry out more systematic studies on screening and designing magnetic materials with nontrivial topological properties.

### I. Two-dimensional magnetic materials

Pioneered by the discovery of graphene,[441] there has been recently a surge of interest on two-dimensional (2D)

materials, due to a vast spectrum of functionalities such as mechanical,[442] electrical,[443] optoelectronic,[444] and superconducting[445] properties and thus immense potential in engineering miniaturized devices. However, the 2D materials with intrinsic long-range magnetic ordering have been missing until the recently confirmed systems like $CrI_3$,[446] $Cr_2Ge_2Te_6$,[447] and $Fe_3GeTe_2$[448] in the monolayer limit. This enables the possibility to fabricate vdW heterostructures based on 2D magnets and hence paves the way to engineer novel spintronic devices. Nevertheless, the field of 2D magnetism is still in its infancy, with many pending problems such as the dimensional crossover of magnetic ordering, tunability, and so on.

According to the Mermin-Wagner theorem,[449] the long-range ordering is strongly suppressed at finite temperature for systems with short-range interactions of continuous symmetry in reduced dimensions, due to the divergent thermal fluctuations. Nevertheless, as proven analytically by Onsager,[450] the 2D Ising model guarantees an ordered phase, protected by a gap in the spin-wave spectra originated from the magnetic anisotropy.[451] In this sense, the rotational invariance of spins can be broken by dipolar interaction, single-ion anisotropy, anisotropic exchange interactions, or external magnetic fields, which will lead to magnetic ordering at finite temperature. The complex magnetic phase diagram is best represented by that of transition metal thiophosphates $MPS_3$ (M = Mn, Fe, and Ni),[452,453] which are of the 2D-Heisenberg nature for $MnPS_3$, 2D-Ising for $FePS_3$, and 2D-XXZ for $NiPS_3$, respectively. As shown in Fig. 8 for few-layer systems of both $FePS_3$ and $MnPS_3$, the transition temperature remains almost the same as that of the corresponding bulk phase. However, for $NiPS_3$, the magnetic ordering is suppressed for the monolayers while few-layer slabs have slightly reduced Néel temperature compared to that of the bulk.[454] This can be attributed to the 2D-XXZ nature of the effective Hamiltonian, where in $NiPS_3$ monolayers it may be possible to realize the BKT topological phase transition.[455] This arises also an interesting question whether there exist novel quantum phases when magnetic ordering is suppressed and how to tune the magnetic phase diagram. Also, it is suspected that the interlayer exchange coupling does not play a significant role for such 2D systems with intralayer AFM ordering.

The interplay of interlayer exchange and magnetic anisotropy leads to more interesting magnetic properties for $CrX_3$ (X = Cl, Br, and I). The $Cr^{3+}$ ions with octahedral environment in such compounds have the $3d^3$ configuration occupying the $t_{2g}$ orbitals in the majority spin channel. The SOC is supposed to be quenched leading to negligible single-ion MAE. Nevertheless, large MAE can be induced by the strong atomic SOC on the I atoms, *i.e.*, intersite SOC, as elaborated in Ref.456,457 As the atomic SOC strength is proportional to the atomic number, the MAE favors in-plane (out-of-plane) magnetization for $CrCl_3$ ($CrBr_3$ and $CrI_3$), respectively. Note that for the bulk phases, the intralayer exchange is FM for all three compounds, while the interlayer exchange is FM for $CrBr_3$ and $CrI_3$, and AFM for $CrCl_3$.[458] In this regard, the most surprising observation for the few-layer systems is that $CrI_3$ bilayers have AFM interlayer coupling[446] with a reduced Néel temperature of 46K in comparison to the bulk value of 61K (Fig. 8). Additionally, the FM/AFM interlayer coupling can be further tuned by pressure, as showed experimentally recently.[459,460] Such a transition from FM to AFM interlayer coupling was suggested to be induced by the stacking fault based on DFT calculations, namely the rhombohedral (monoclinic) stacking favors FM (AFM) interlayer exchange.[461] It is noted that $CrX_3$ has the structural phase transition in the bulk phases from monoclinic to rhombohedral around 200 K, e.g., 220K for $CrI_3$.[462] Thus, a very interesting question is why there is a structural phase transition from the bulk rhombohedral phase to the monoclinic phase in bilayers. A recent experiment suggests that such a stacking fault is induced during exfoliation, where it is observed that the interlayer exchange is enhanced by one order of magnitude for $CrCl_3$ bilayers.[463] It is noted that it is still not clear how the easy-plane XY magnetic ordering in $CrCl_3$ would behavior in the monolayer regime. Therefore, for such 2D magnets, an intriguing question to explore is the dimensional crossover, *i.e.*, how the magnetic ordering changes few-layer and monolayer cases in comparison to the bulk phases, as we summarized in Fig. 8 for the most representative 2D magnets.

The key to understand the magnetic behavior of such 2D magnets is to construct a reliable spin Hamiltonian based on DFT calculations. Different ways to evaluate the interatomic exchange parameters have been scrutinized in Ref.464 for $CrCl_3$ and $CrI_3$. It is pointed out that the energy mapping method, which has been widely used, leads to only semi-quantitative estimation. Whereas the linear response theory including the ligand states give more reasonable results, ideally carried out in a self-consistently way. Particularly, exchange parameters beyond the nearest neighbors should be considered, as indicated by fitting the inelastic neutron scattering on $FePS_3$.[465] Moreover, for the interlayer exchange coupling, DFT calculations predict AFM coupling while the ground state is FM for $Cr_2Si_2Te_6$,[466] which may be due to the dipole-dipole interaction which is missing in non-relativistic DFT calculations.[467] Note that the interlayer exchange coupling is very sensitive to the distance, stacking, gating, and probably twisting,[468] which can be used to engineer the spin Hamiltonian which should be evaluated quantitatively to catch the trend. Furthermore, as the most of the 2D magnets have the honeycomb lattice with edge-sharing octahedra like the layered $Na_2IrO_3$ and $\alpha$-$RuCl_3$ which are both good candidates to realize the quantum spin liquid states,[469] the Kitaev physics becomes relevant, *e.g.*, a comparative investigation on $CrI_3$ and $Cr_2Ge_2Te_6$ demonstrated that there does exist significant Kitaev interactions in both compounds.[470] Thus, for 2D magnets, the full 3×3 exchange matrix as speci-fied in Eq. (IV C) has to be considered beyond the nearest neighbors, in order to understand the magnetic ordering and emergent properties. This will enable us to engineer the parameter space to achieve the desired phases, *e.g.*, to obtain 2D magnetism with critical temperature higher than the room temperature. For instance, the antisymmetric DMI can stabilize skyrmions in 2D magnets, as observed in recent experiments on $Fe_3GeTe_2$ nanolayers.[471,472]

Due to the spatial confinement and reduced dielectric screening, the Coulomb interaction is supposed to be stronger in 2D materials, which bring forth the electronic correlation physics. Taking $Cr_2Ge_2Te_6$ as an example, recent ARPES measurements demonstrate that the band gap is about 0.38 eV at 50K,[473] in comparison to 0.2 eV at 150K,[474] while the band gap is only about 0.15 eV in the GGA+U calculations with U = 2.0 eV which reproduce the MAE.[475] As both the magnetic ordering and electronic correlations can open up finite band gaps, recent DFT+DMFT calculations reveal that the band gap in $Cr_2Si_2Te_6$ is mostly originated from the electronic correlations.[476] Additionally, both the *d*-*d* transitions and molecular ligand states are crucial for the helical photoluminescence of $CrI_3$ monolayers.[477] Moreover, it is also demonstrated that the ferromagnetic phase coexists with the Kondo lattice behavior in $Fe_3GeTe_2$.[478] The correlated nature of the electronic states in 2D magnets calls for further experimental investigation and detailed theoretical modelling. 2D magnetic materials provide also an interesting playground to study the interplay between electronic correlations and magnetism, which is in competition with other emergent phases.

To tailor the 2D magnetism, voltage control of magnetism has many advantages over the mainstream methods for magnetic thin films such as externally magnetic fields and electric currents, because they suffer from poor energy efficiency, slow operating speed, and appalling size compactness. For biased $CrI_3$ bilayers, external electric fields can be applied to switch between the FM and AFM interlayer coupling.[479,480] Such a gating effect can be further enhanced by sandwiching dielectric BN layers between the electrode and the $CrI_3$ layers.[481] Moreover, ionic gating on $Fe_3GeTe_2$ monolayers can boost their Curie temperature up to the room temperature, probably due to enhanced MAE but with quite irregular dependence with respect to the gating voltage.[482] Interesting questions are to understand the variation in the spin Hamiltonian and thus the Curie temperature in such materials under finite electric fields. Additionally, strain has also been applied to tailor 2D magnets. For instance, it is demonstrated that 2% biaxial compressive strain leads to a magnetic state transition from AFM to FM for $FePS_3$ monolayers,[483] and 1.8% tensile strain changes the FM ground state into AFM for $CrI_3$ monolayers.[484] Moreover, strain can also induce significant modification of the MAE, e.g., 4% tensile strain results in a 73% increase of MAE for $Fe_3GeTe_2$ with a monotonous dependence in about 4% range.[485] Thus, it is suspected that biaxial strain can be applied to tune the magnetic properties of vdW magets effectively.

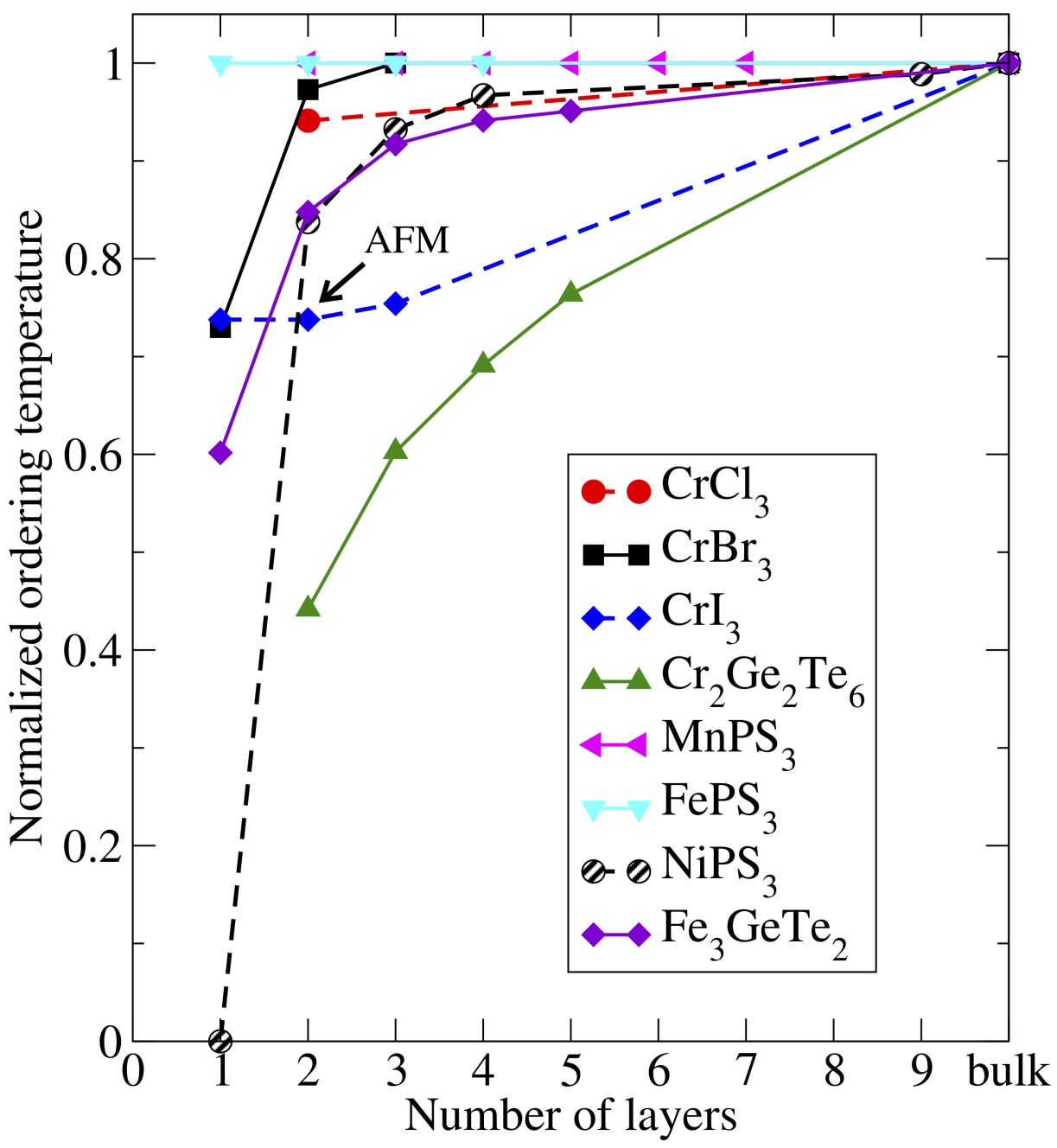

FIG. 8: Dependence of the magnetic ordering temperature with respect to the number of layers for typical 2D magnets. The values are normalized using the corresponding bulk Curie/Néel temperature, e.g., $CrCl_3$ ($T_C$ = 17K),[458] $CrBr_3$ ($T_C$ = 37K),[458] $CrI_3$ ($T_C$ = 61K),[446] $Cr_2Ge_2Te_6$ ($T_C$ = 68K),[447] $MnPS_3$ (TN = 78K),[464] $FePS_3$ (TN = 118K),[465] $NiPS_3$ ($T_N$ = 155K),[454] and $Fe_3GeTe_2$ ($T_C$ = 207K).[448] Lines are guide for the eyes. The figure is adapted from Ref. 486 with updated data. Copyright requested.

2D magnets exhibit a wide spectrum of functionalities. For instance, tunnelling magnetoresistance in heterostructures of CrI3 has been observed in several set-ups incorporating $CrI_3$ multilayers,[487–489] with the magnetoresistance ratio as large as 1,000,000%.[490] Such a high ratio can be attributed to perfect interfaces for the vdW heterostructures, which is challenging to achieve for conventional heterostructures of magnetic metals and insulators such as Fe/MgO. The tunnelling can be further tuned via gating,[491] leading to implementation of spin tunnel field-effect transistors.[492] Moreover, metallic 2D magnets displays also fascinating spintronic properties, *e.g.*, spin valve has been implemented based on the sandwich $Fe_3GeTe_2$/BN/$Fe_3GeTe_2$ geometry, where a tunnelling magnetoresistance of 160% with 66% polarization has been achieved.[493]

Such transport properties are based on the charge and spin degrees of freedom of electrons, whereas emergent degree of freedom such as valley has attracted also intensive attention, leading to valleytronics.[494] As most of the compounds discussed above have the honeycomb lattice, the valley-degeneracy in the momentum space will be lifted when the inversion symmetry is broken, resulting in valley-dependent (opto-)electronic properties. Moreover, considering SOC causes the spin-valley coupling,[495] *i.e.*, valley-dependent optical selection rules become also spin-dependent, which can be activated by circularly polarized lights. Such phenomena can be realized in the vdW heterostructures such as the $CrI_3/WSe_2$ heterostructures[496] via the proximity effect or in AFM 2D magnets such as $MnPS_3$.[497]

From the materials perspective, 2D materials have been extensively investigated both experimentally and theoretically. There exist several 2D material databases[498–501] based on HTP DFT calculations, leading to predictions of many interesting 2D magnets.[502] Nevertheless, it is still a fast developing field, *e.g.*, the Curie temperature of 2D FM compounds has been pushed up to the room temperature in $VTe_2$[503] and $CrTe_2$.[504] It is noted that the DFT predictions should go hand-in-hand with detailed experimental investigation. For instance, $VSe_2$ monolayers are predicted to be a room temperature magnet[505] supported by a follow-up experiment.[506] However, recent experiments reveal that the occurrence of the charge density wave phase will prohibit the magnetic ordering.[507,508] To go beyond the known prototypes to design new 2D materials, one can either predict more exfoliable bulk compounds with layered structure or laminated compounds where etching can be applied. For instance, MXene is a new class of 2D materials which can be obtained from the MAX compounds,[509] where the HTP calculations can be helpful. It is particularly interesting to explore such systems with $4d/5d$ TM or RE elements, where the interplay of electronic correlations, exchange coupling, and SOC will leads to more intriguing properties, such as recently reported $MoCl_5$[510] and $EuGe_2$.[511] Last but not least, vdW materials can also be obtained using the bottom-up approach such as molecular beam epitaxy down to the monolayer limit, such as $MnSe_2$ on GaSe substrates.[512] This broadens significantly the materials phase space to those non-cleavable or metastable vdW materials and provides a straightforward way to in-situ engineer the vdW heterostructures.

Like the 3D magnetic materials, 2D magnets displays comparable properties but with a salient advantage that they can form van der Waals (vdW) heterostructures without enforcing lattice matching.[513] This leads to a wide range of combinations to further tailor their properties, *e.g.*, the emergent twistronics[514] exampled by the superconductivity in magic angle bilayer graphene.[515] In comparison to the interfaces in conventional magnetic heterostructures, which are impeded by dangling bonds, chemical diffusion, and all possible intrinsic/extrinsic defects, vdW heterostructures with well-defined interfacial atomic structures are optimal for further engineering of spintronic devices.[516] Nevertheless, to the best of our knowledge, there is no 2D magnet showing magnetic ordering at room temperature, which is a challenge among many as discussed above for future HTP design.

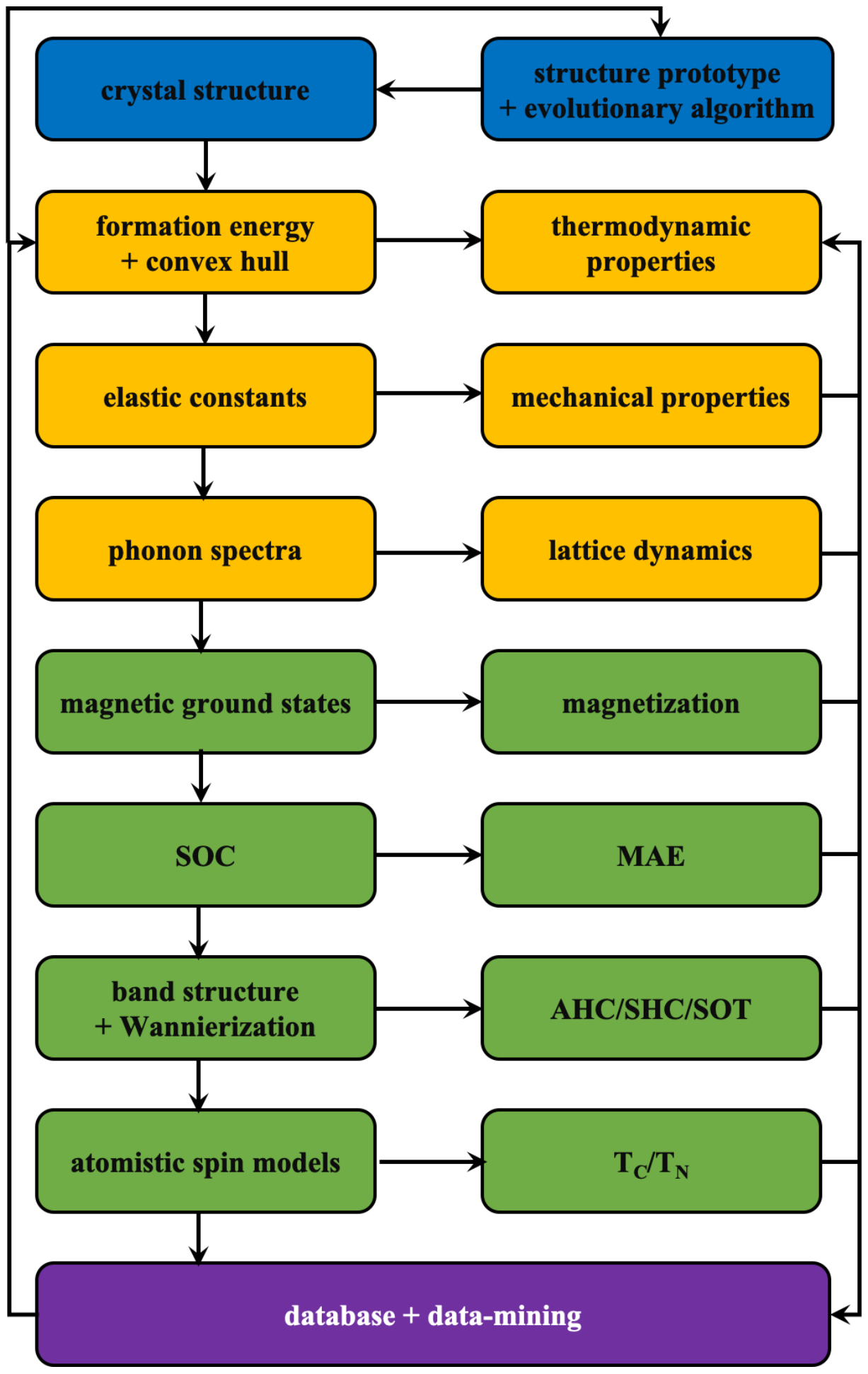

FIG. 9: A typical HTP workflow to design magnetic materials. Blue, yellow, green, and purple blocks denote the processes of crystal structure identification, evaluation of stabilities, characterization of magnetic properties, and database curation, respectively.

## V. CASE STUDIES

### A. High-throughput workflows

Fig. 9 displays a generic workflow to perform HTP screening on magnetic materials, with four processes highlighted in different colors. That is, HTP screening can be performed in four essential steps:

1. crystal structure identification. As DFT calculations need only the chemical composition and crystal structure as inputs, HTP calculations mostly start with such information which can be obtained via (1) databases compiling known compounds, (2) substitution based on crystal structure prototypes, and (3) crystal structure prediction using evolutional algorithms. There have been a few databases such as ICSD https://icsd.fiz-karlsruhe.de/index.xhtml

and COD http://www.crystallography.net/cod/, with the crystal structures for experimentally synthesized compounds and theoretically predicted crystal structures. Moreover, typical structure prototypes have been identified,[54,55]which offer a good starting point to perform chemical substitutions.[517] The crystal structures can also be generated based on evolutionary algorithms using USPEX[518] and CALYPSO.[60]

2. evaluation of stabilities. This is usually done in a funnel-like way, as thermodynamic, mechanical, and dynamical stabilities should be systematically addressed, as detailed in Sect. IV A. We want to emphasise that the evaluation of convex hull should be performed with respect to as many competing phases as possible, thus the corresponding DFT calculations are better done using consistent parameters as cases in existing databases such as Materials Project,[8] AFLOW,[9] and OQMD.[11]

3. characterization of magnetic properties. In Fig. 9 we list several fundamental magnetic properties, such as magnetization, MAE, $T_C/T_N$ and transport properties (please refer to the previous sections for detailed discussions), which can be selectively evaluated depending on the target applications. It is worthy pointing out that a systematic evaluation of the thermodynamic properties for magnetic materials dictates accurate Gibbs free energies with significant magnetic contributions,[519] as illustrated for the Fe-N systems.[520] Thus, proper evaluation of the magnetization and magnetic ordering temperature is required.

4. database curation. Ideally, a data infrastructure is needed in order to obtain, store, analyze, and share the DFT calculations. Non-SQL databases (*e.g.*, MongoDB) are mostly used because of their flexibility with the data structure, which can be adjusted for compounds with various complexity and add-on properties evaluated in an asynchronized way. Moreover, the data should be findable, accessible, interoperable, and reusable (FAIR),[521] so that they can be shared with the community with approved provenance. Last but not least, with such database constructed, data mining using machine learning techniques can be performed to further accelerate the previous three steps and develop statistical insights on the results, as discussed in Sect. VI B.

It is straightforward to set up such HTP workflows, either based on integrated platforms or starting from scratch, as many DFT codes have Python interfaces.[522] In this regard, the atom simulation environment[523] is a convenient tool, which has been interfaced to more than 30 software packages performing DFT and molecular simulations. Nevertheless, the following two aspects deserve meticulosity, which essentially distinguish HTP from conventional DFT calculations. The workflow should be automated, so that the computational tasks can be properly distributed, monitored, and managed. For instance, a typical practice is to generate input files for thousands of compositions, optimize the crystal structures, and assess the thermodynamic stability. The corresponding workflows can be implemented using stand-alone Python packages like fireworks,[524] or using the integrated platforms such as AiiDa[12] and Atomate.[13] Another critical point is job management and error handling. Ideally, a stand-by thread shall monitor the submitted jobs, check the finished ones, and (re-)submit further jobs with proper error (due to either hardware or software issues) recovery. In this regard, custodian(https://materialsproject.github.io/custodian/) is a good option, with integrated error handling for the VASP, NwChem, and QChem codes. On the other hand, the set-up and maintenance of such job management depends on the computational environment, and also expertise with the usage of scientific softwares.

### B. Heusler compounds

Heusler compounds form an intriguing class of intermetallic systems possessing a vast variety of physical properties such as HM, MSME, TI, superconductivity, and thermoelectricity (please refer to Ref. 525 for a comprehensive review). Such versatility is originated from the tunable metallic and insulating nature of the electronic structure due to the flexibility in the chemical composition. As shown in Fig. 10, there are two main groups of Heusler compounds, namely, the half Heusler with the chemical composition XYZ and the full-Heusler $X_2YZ$, where X and Y are usually transition metal or RE elements, and Z being a main group element. For both half Heusler and full Heusler systems, the crystal structures can be considered to be with a skeleton of the ZnS-type (zinc blende) formed by covalent bonding between X and Z elements, and the Y cations will occupy the octahedral sites and the vacancies. This gives rise to the flexibility to tune the electronic structure by playing with the X, Y, and Z elements.[525]

The full Heuslers $X_2YZ$ can end up with either the regular ($L2_1$-type) or the inverse ($X_\alpha$-type) Heusler structure, as shown in Fig. 10. It is noted that the both the regular and inverse full Heusler crystal structures are derived from the closely packed fcc structure, which is one of the most preferred structures for ternary intermetallic systems.[526] Depending on the relative electronegativity of the X and Y elements, the empirical Burch's rule states that the inverse Heusler structure is preferred if the valence of Y is larger than that of X, *e.g.*, Y is to the right of X if they are in the same row of the periodic table.[527] Several HTP calculations have been done to assess the Burch's rule, such as in Sc-[528] and Pd-based[529,530] Heuslers. Additionally, the nonmagnetic $L2_1$ Heusler compounds as high-strength alloys[531] and thermoelectric[532] materials have been studied. For magnetic properties, Sanvito *et al.*[57] performed HTP calculations on 236,115 $X_2YZ$ compounds and found that 35,602 compounds are stable based on the formation energy, where 6,778 compounds are magnetic. Unfortunately, the relative stability with respect to competing binary and ternary phases is expensive thus demanding to accomplish. In a recent work,[533] Ma *et al.* carried out DFT calculations on a small set of 405 inverse Heusler compounds, but with the relative stability to the regular Heusler and the convex hull considered systematically, including also a few possible magnetic configurations.[533]Ten HMs have been identified, which can be explored for spintronic applications.

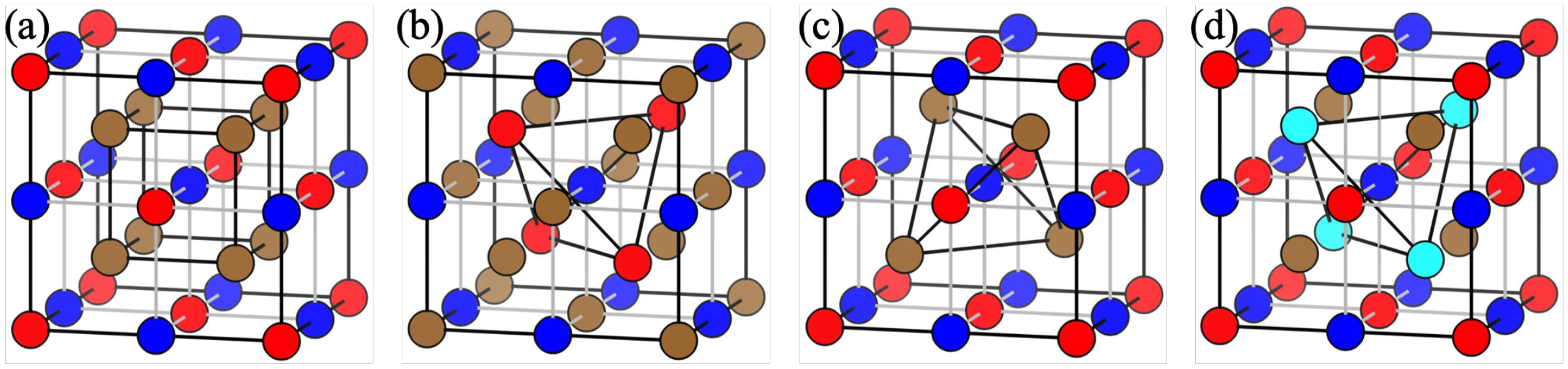

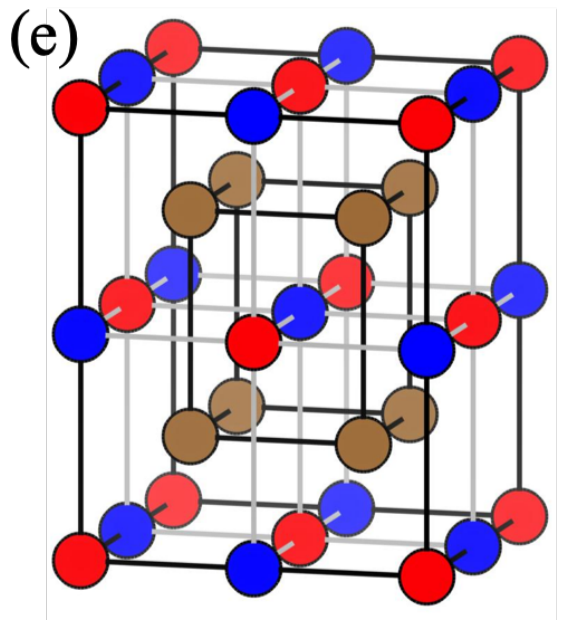

| Wyckoff position F$\bar{4}$3m | (regular) full Heusler | Inverse (full) Heusler | Half Heusler | Quaternary Heusler |
|---|---|---|---|---|
| 4a (0, 0, 0) | Y | X | Y | Y |
| 4b (1/2, 1/2, 1/2) | Z | Z | Z | Z |
| 4c (1/4, 1/4, 1/4) | X | Y | - | X′ |
| 4d (3/4, 3/4, 3/4) | X | X | X | X |

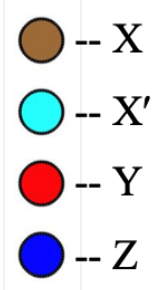

FIG. 10: Illustration of the Heusler crystal structures for (a) (regular) full Heusler $X_2YZ$, (b) inverse (full) Heusler $X_2YZ$, (c) half Heusler XYZ, (d) quaternary Heusler XX′YZ, and (e) tetragonal (full) Heusler $X_2YZ$. The conventional unit cell of 16 atoms are sketched, using the Wyckoff positions of the F$\bar{4}$3m to give a general description for the cubic (a-d) cases.

Focusing on the regular Heusler systems, Balluff *et al.*[534] performed systematic evaluation of the convex hull based on the available data in AFLOW and it is found that the number of magnetic Heuslers will be reduced from 5,000 to 291. That is, the convex hull construction will reduced the number of stable compounds by one order of magnitude, consistent with our observation on the magnetic antiperovskites.[46] Therefore, although the database of competing phases might not be complete, it is highly recommended to evaluate the thermodynamic stability including both the formation energy and convex hull, if not the mechanical and dynamical stabilities. Furthermore, to identify the magnetic ground states, Balluff *et al.* carried out calculations on two antiferromagnetic configurations and obtained 70 compounds with AFM ground states. The Néel temperature of such compounds are then computed using the Monte Carlo method with exchange parameters calculated explicitly using the SPRKKR (https://ebert.cup.uni-muenchen.de) code, resulting in 21 AFM Heusler compounds with Néel temperature higher than 290 K. Such compounds are ready to be explored for applications on AFM spintronics.

Interestingly, for both regular and inverse Heusler compounds, tetragonal distortions can take place driven by the band Jahn-Teller effect associated with high DOS at the Fermi energy.[535] HTP calculations reveal that 62% of the 286 Heusler compounds investigated prefer the tetragonal phase, due to the van Hove singularities around the Fermi energy.[536] Such tetragonal Heusler compounds exhibit large MAE, *e.g.*, 5.19 MJ/m$^3$ for $Fe_2PtGe$ and 1.09 MJ/m$^3$ for $Fe_2NiSn$, making them interesting candidates as permanent magnets and STT-MRAM materials.[537] To search for potential materials for STT applications, Al-, Ga-, and Sn-based inverse Heusler compounds in both cubic and tetragonal structures have been investigated, aiming at optimizing the spin polarization and Gilbert damping for materials with perpendicular magnetic anisotropy.[538] Additionally, such tetragonal Heusler compounds can also be used to engineer magnetic heterostructures with enhanced TMR effects.[539]

Significant MAE can also be obtained by inducing tetragonal distortions on the cubic Heusler compounds. For instance, the MAE of $Ni_2YZ$ compounds can be as large as 1 MJ/m$^3$ by imposing $c/a \neq 1$.[540] To stabilize such tetragonal distortions, we examined the effect of light elements (H, B, C, and N) as interstitial dopants into the Heuslers and found that tetragonal distortions can be universally stabilized due to the anisotropic crystalline environments for the interstitials preferentially occupying the octahedral center.[541] This leads to an effec-tive way to design RE-free permanent magnets. Two pending problems though are the solubility and possible disordered distribution of the light interstitials in the Heusler structures. Nevertheless, it is demonstrated that for the Fe-C alloys, due to interplay of anharmonicity and segregation, collective ordering is preferred,[542] entailing further studies on the behavior of interstitials in the Heusler compounds. We note that the structural phase transition between the cubic and tetragonal phases in $Ni_2MnGa$ induces the MSME and in $Ni_2MnX$ (X = In, Al, and Sn) the MCE. In this regard, there are many more compounds with such martensitic transitions which can host MSME and MCE, awaiting further theoretical and experimental investigations.

To go beyond the *p*-*d* covalent bonding between the X and Z atoms which stabilizes the Heusler structures, Wei *et al.* first explored the *d*-*d* hybridization in the Ni-Mn-Ti systems and observed that there exist stable phases with significant MSME, which offers also the possibility to improve the mechanical properties of the resulting Heusler systems in comparison to the main-group-element based conventional cases.[543] Considering only the cubic phases, follow-up HTP calculations revealed that 248 compounds are thermodynamically stable out of 36,540 prototypes and 22 of them have magnetic ground states compatible with the primitive unit cell.[57] The predictions are validated by successful synthesis of $Co_2MnTi$ and $Mn_2PtPd$ which adopts the tetragonal structure with space group I4/mmm (Fig. 10).[57] An interesting question is whether there exist promising candidates with enhanced caloric performance beyond the known Ni-Mn-Ti case.[544] There have been several studies focusing on Zn-,[545] V-,[546] and Cd-based[547] Heuslers, where possible martensitic transformation is assessed based on the Bain paths. We note that in order to get enhanced MCE, the martensitic phase transition temperature is better aligned within the Curie temperature window.[548] Therefore, to predict novel all-d-metal Heuslers with magneto-structural functionalities, more systematic HTP characterization on the thermodynamic and magnetic properties for both cubic and tetragonal phases is required.

The flexibility in the chemical composition for the intermetallic Heusler compounds can be further explored by substituting a fourth element or vacancy in $X_2YZ$, leading to quaternary XX′YZ and half XYZ Heusler (Fig. 10). Following the empirical rule, we performed HTP calculations on magnetic quaternary Heusler compounds with 21, 26, and 28 valence electrons to search for spin gapless semiconductors (SGSs).[549] Considering both structural polymorphs by shuffling atomic positions and magnetic ordering compatible with the primitive cell, we identified 70 unreported candidates covering all four types of SGSs out of 12,000 chemical compositions, with all the 22 experimentally known cases validated. Such SGSs are promising for spintronic applications, as they display significant anisotropic magnetoresistance, and tunable AHC and tunnelling magnetoresistance.[549,550] We note that the semimetallic phase can also be found in quaternary Heuslers with the other numbers of electrons,[551] which is interesting for future studies.

Regarding the half Heusler compounds, there have been extensive HTP calculations to screen for nonmagnetic systems as thermoelectric materials,[552,553] where the formation of defects has also been addressed in a HTP manner.[554] Ma *et al.* carried out HTP calculations on 378 half Heuslers and identified 26 semiconductors, 45 HMs, and 34 nearly HMs, which are thermodynamically stable.[83] Another recent work investigated the alkaline element based half Heuslers and found 28 ferromagnetic materials out of 90 compositions.[555]

Importantly, for half Heusler materials, it is not enough to consider only the binary competing phases, but also possible ternary compounds as there are many crystal structures which can be stabilized based on the 18-electron rule for the 1:1:1 composition.[556] For instance, the so-called hexagonal Heusler compounds are a class of materials which can host topological insulator[557] and (anti-)ferroelectric phases.[558,559] Particularly, the magnetic counterparts hexagonal Heuslers host martensitic transitions, making them promising for MCE applications.[560] This leads to two recent work where the composition of the hexagonal Heusler compounds is optimized to further improve the MCE,[561,562] whereas more systematic HTP calculations are still missing. Additionally, it is demonstrated that the half Heuslers with the $F\bar{4}3m$ structure can be distorted into either $P6_3mc$ or Pnma structures, giving rise to an interesting question whether magnetic materials with significant MCE can be identified based on HTP screening. Last but not least, like the quaternary Heuslers, the flexibility in composition can also be realized in the half Heusler structure. For instance, 131 quaternary double half Heusler compounds are predicted to be stable where $Ti_2FeNiSb_2$ has been experimentally synthesized showing low thermal conductivity as predicted. This paves the way to explore the tunable electronic and magnetic properties of half Heusler compounds.

One particularly interesting subject is to screen for Heuslers with transport properties of the topological origin. For instance, motivated by the recent discovered Weyl semimetal $Co_2MnGa$,[359] an enlightening work evaluated the topological transport properties (*e.g.*, AHC, ANC, and MOKE which obey the same symmetry rules) of 255 cubic (both regular and inverse) Heusler compounds in a HTP way.[563] By comparing the results for full and inverse Heuslers systems, it is observed that the mirror symmetry present in the full but not the inverse Heusler compounds plays an essential role to induce significant linear response properties. It is noted that only one FM (AFM) configuration is considered for the full (inverse) Heusler compounds, which might not be enough as the electronic structure and the derived physical properties are subjected to the changes in the magnetic configurations.

Overall, it is obvious that Heusler compounds are a class of multifunctional magnetic materials, but the cur-rent HTP calculations are still far away from being complete. One additional issue is how to properly treat the chemical and magnetic disorders in such compounds in order to predict the electronic and thermodynamic properties. For instance, the MCE takes place only for Mn-rich Ni-Mn-Ti,[543] which has not yet been explained based on DFT calculations. We believe substitutions and chemically disordered systems deserve more systematic treatments in the future HTP studies, particularly for Heusler alloys.

### C. Permanent magnets

As discussed above, the main tasks for HTP screening of permanent magnets are to evaluate the intrinsic magnetic properties for the known and predicted phases, including both RE-free and RE-based systems. A recent work[564] focuses on the RE-free cases, where starting from 10,000 compounds containing Cr, Mn, Fe, Co, and Ni available in the ICSD database, step-by-step screening is carried out based on the chemical composition, $M_s$, crystal structure, magnetic ground state, and MAE. Three compounds, namely, $Pt_2FeNi$, $Pt_2FeCu$, and $W_2FeB_2$ are finally recommended. It is noted that the magnetic ground state for the compounds during the screening is determined by the literature survey, which might be labor intensive and definitely cannot work for "unreported" cases.

In addition to HTP characterization of the known phases and prediction of novel compounds, tailoring the existing phases on the boundary of being permanent magnets is also promising, which can be done via substitutional and interstitial doping. For instance, theoretical predictions suggest that Ir/Re doping can enhanced the MAE of $(Fe_{1-x}Co_x)_2B$, which is confirmed by experiments.[565] Nevertheless, cautions are deserved when evaluating the MAE for the doped compounds using the supercell method, where the symmetry of the solution phases may be broken thus leading to erroneous predictions on MAE. Special techniques such as VCA and CPA are needed to obtain MAE as done in Ref. 565, which are not generally available in the main stream codes. Furthermore, the interstitial atoms such as H, B, C, and N can also be incorporated into the existing intermetallic phases, giving rise to structural distortions and thus significant MAE as demonstrated for the FeCo alloys.[566,567] Following this line, we have performed HTP calculations to investigate the effects of interstitials on the intermetallic compounds of both the $Cu_3Au$[82] and Heusler[541] types of crystal structures, and identified quite a few candidates interesting for further exploration.

Turning now to the RE-based permanent magnets, due to the challenge to treat the correlated $4f$-electrons, approximations are usually made. For instance, in a series of work, the tight-binding linear muffin-tin orbital method in the atom sphere approximation has been applied to screen the 1-5/2-17/2-14-1 type,[568] 1-12 type,[569,570] 1-11-X type,[173] and the 1-13-X type[571] classes of materials, where the $4f$-electrons are considered in the spin-polarized core approximation. One apparent problem is that the magnetization and MAE cannot be obtained consistently based on such a method. The Staunton group has developed a consistent theoretical framework to evaluate both the magnetization and MAE,[572] which is interesting if further HTP calculations can be performed. Additionally, another common problem to access the thermodynamic stability of the RE-based compounds with most DFT codes on the market. Nevertheless, the trend with respect to chemical compositions can be obtained, which can be validated with further experiments. Last but not least, the spin moments of the RE elements is usually aligned antiparallel to the magnetic moment of the transition metal atoms, which is usually not properly treated. For such materials, not only the single ion MAE but also inter-sublattice exchange coupling should be included in order to compare with experimental measurements.[573]

Therefore, due to the challenges such as the correlated nature of $4f$-electrons and its interplay with SOC, HTP design of permanent magnets is still a subject requiring further improvement on the computing methodology. As such materials are applied in big volume, it might be strategic to consider firstly the criticality of the constitute elements, *e.g.*, the high cost and supply risk on Co, Nd, Dy, and Tb. In this regard, compounds based on Fe and Mn combined with cheap elements in the periodic table should be considered with high priority. In addition, all three intrinsic properties ($M_s$, MAE, and $T_C$) should be optimized which is numerically expensive and thus not usually done. In this regard, a consistent figure of merit should be adopted, *e.g.*, the suggestion by Coey to define and optimize the dimensionless magnetic hardness parameter $\kappa = \left(\frac{K_1}{\mu_0 M_s^2}\right)^{1/2} > 1$, which should be implemented for future HTP screening.[29]

### D. Magnetocaloric materials

As discussed in Sect. IV F, MCE is driven by the interplay of various degrees of freedom leading to challenges in performing HTP design of novel MCE materials. For instance, rigorous evaluation of the Gibbs free energies including the lattice, spin, and electronic degrees of freedom is a numerically expensive task, not to mention the complex nature of magnetism and phase transition. Thus, it is of great impetus to establish a computational "proxy" which correlates the MCE performance with quantities easily accessible via DFT calculations.

Following the concept that MCE is significant when the magneto-structural phase transitions occur, Bocarsly *et al.* proposed a proxy in terms of the magnetic deformation $\Sigma_M = \frac{1}{3}(\eta_1^2 + \eta_2^2 + \eta_3^3)^{1/2} \times 100$ and $\boldsymbol{\eta} = \frac{1}{2}(\mathbf{P}^T\mathbf{P} - \mathbf{I})$ where $\mathbf{P} = \mathbf{A}_{\text{nonmag}}^{-1} \cdot \mathbf{A}_{\text{mag}}$ with $\mathbf{A}_{\text{nonmag}}$ and $\mathbf{A}_{\text{mag}}$ being the lattice constants of the nonmagnetic and mag-netic unit cells.[574] Assuming the high temperature paramagnetic phase can be described with the nonmagnetic solution at 0K, the magnetic deformation $\Sigma_M$ measures "*the degree to which structural and magnetic degrees of freedom are coupled in a material*". In fact, there is a universal correlation between $\Delta S$ and $\Sigma_M$, no matter whether MCE is driven by FOPTs or not. Nevertheless, there is no direct scaling between $\Delta S$ and $\Sigma_M$ and it is suggested that $\Sigma_M > 1.5\%$ is a reasonable cutoff to select the promising compounds. Further screening on the known FM materials reveals 30 compounds out of 134 systems as good candidates, where one of them MnCoP is validated by experimental measurements showing $\Delta S = -3.1 J/kg/K$ under an applied field of 2T. Recently, this proxy has been successfully applied to evaluate the MCE behavior of two solid solutions $Mn(Co_{1-x}Fe_x)Ge$ and $(Mn_{1-y}Ni_y)CoGe$, where the predicted optimal compositions x=0.2 and y=0.1 are in good agreement with the experiments.[562] It is interesting that such consistency is obtained by evaluating the corresponding quantities via configurational average over supercells with Boltzmann weights, instead of the SQS method. One point to be verified is whether the nonmagnetic instead of the paramagnetic configurations are good enough. Our preliminary results on a serious of APVs with noncollinear magnetic ground states indicate that the negative thermal expansion associated with the magneto-structural transition and hence MCE is overestimated using the magnetic deformation between FM and nonmagnetic states.

A more systematic and computationally involved workflow is the CaloriCool approach,[237] based on two-step screening. It is suggested that the metallic alloys and intermetallic compounds are more promising candidates than the oxides, which "*suffer from intrinsically low isentropic (a.k.a. adiabatic) temperature changes due to large molar lattice specific heat*". The first step fast screening is based on the phase diagrams and crystal structures from the literature and known databases, which results in compounds with the same chemical composition but different crystal structures. In the second step, the physical properties including mechanical, electronic, thermodynamic, kinetic properties will be evaluated, c ombining DFT with multi-scale and thermodynamic methods. For instance, the thermodynamic properties can be evaluated following the methods proposed in Ref. 575, which is foreseeably expensive and better done one after another. Also, the success of the approach depends significantly on the database used for the first step screening. Nevertheless, it is proposed based on such design principles that there should be a solution phase $Zr_{1-x}Yb_xMn_6Sn_6$ with significant MCE due to the fact that the pristine $ZrMn_6Sn_6$ and $YbMn_6Sn_6$ are with AFM and FM ground states with critical temperatures being 580K and 300K, respectively. This is very similar to the morphotropic phase boundary concept for ferroelectric materials, where the piezoelectric response associated with first order ferroelectric phase transition can be greatly enhanced at the critical compositions.[576]

## E. Topological materials

As discussed in Sect. IV H, insulators and semimetals of nontrivial nature are an emergent class of materials from both fundamental physics and practical applications points of view. To design such materials, symmetry plays an essential role as explained in detail in Sect. IV H for a few specific compounds. Particularly for magnetic materials, the occurrence of QAHI and Weyl semimetals is *not* constrained to compounds of specific symmetries, whereas the AFM TIs entail a product symmetry $\mathcal{S} = \mathcal{P}\Theta$ which can give rise to the demanded Kramers degeneracy. It is noted that the symmetry argument applies to the nonmagnetic topological materials as well, where based on topological quantum chemistry[577] and filling constraints[578,579] HTP characterization of nonmagnetic materials have been systematically performed.[580–583] Such screening can also be performed in a more brute-force way by calculating the surface states[584] and spin-orbit spillage.[585,586]

A very fascinating work done recently is to perform HTP screening of AFM topological materials based on the magnetic topological quantum chemistry,[587] as an extension to the topological quantum chemistry.[577] For this approach, the irreducible co-representations of the occupied states are evaluated at the high symmetry points in the BZ, and the so-called compatibility relations are then taken to judge whether the compounds are topologically trivial or not.[588] In this way, six categories of band structures can be defined, *e.g.* band representations, enforced semimetal with Fermi degeneracy, enforced semimetal, Smith-index semimetal, strong TI, and fragile TI, where only band representations are topologically trivial. Out of 403 well converged cases based on DFT calculations of 707 compounds with experimentally available AFM magnetic structure collected in the MAGNDATA database,[589] it is observed that about 130 ($\approx$ 32%) showing nontrivial topological features, where NpBi, $CaFe_2As_2$, NpSe, $CeCo_2P_2$, $MnGeO_3$, and $Mn_3ZnC$ are the most promising cases. We note that one uncertainty is the U value which is essential to get the correct band structure, and more sophisticated methods such as DFT+DMFT might be needed to properly account for the correlations effect.[184]

## F. 2D magnets

The HTP screening of 2D magnetic materials has been initiated by examining firstly the bulk materials that are held together by the van der Waals interaction, which can then be possibly obtained by chemical/mechanical exfoliation or deposition. Starting from the inorganic compounds in the ICSD database,[590] two recent publications tried to identify 2D materials by evaluating the packing ratio, which is defined as the ratio of the covalent volume and the total volume of the unit cell.[591,592] This leads to about 90 compounds which can be obtained in the 2D form, where about 10 compounds are found to be magnetic.[592] The screening criteria have then be generalized using the so-called topological-scaling algorithm, which classifies the atoms into bonded clusters and further the dimensionality.[498] Combined with the DFT evaluation of the exfoliation energy, 2D materials can be identified from the layered solids. This algorithm has been applied to compounds in the Materials Project database, where 826 distinct 2D materials are obtained. Interestingly, 128 of them are with finite magnetic moments $>$ 1.0 $\mu_B$/u.c., and 30 of them showing HM behavior.[498] Extended calculations starting from the compounds in the combined ICSD and COD databases identified 1825 possible 2D materials, and 58 magnetic monolayers have been found out of 258 most promising cases.[499] Similarly, by evaluating the number of "covalently connected atoms particularly the ratio in supercell and primitive cells, 45 compounds are identified to be of layered structures out of 3688 systems containing one of (V, Cr, Mn, Fe, Co, Ni) in the ICSD database, leading to 15 magnetic 2D materials.[593]

As discussed in Sect. IV I, the occurrence of 2D magnetism is a tricky problem, particularly the magnetic ordering temperature which is driven by the interplay of magnetic anisotropy and exchange parameters. Based on the computational 2D materials database (C2DB),[500] HTP calculations are performed to obtain the exchange parameters, MAE, and the critical temperature for 550 2D materials, and it is found that there are about 150 (50) FM (AFM) compounds being stable.[502] Importantly, the critical temperatures for such compounds have a strong dependence on the values of U, indicating that further experimental validation is indispensable. Nevertheless, the characterization of the other properties for 2D materials in the HTP manner is still limited. One exception is the HTP screening for QAHC with in-plane magnetization,[594] where the prototype LaCl might not be a stable 2D compound. It is noted that there have been a big number of predictions on 2D magnetic and nonmagnetic materials hosting exotic properties but the feasibility to obtain such compounds has not been systematically addressed. Therefore, we suspect that consistent evaluations of both the stability and physical properties are still missing, particularly in the HTP manner which can guide and get validated by future experiments.

Despite the existing problems with stability for 2D (magnetic) materials, a system workflow to characterize their properties has been demonstrated in a recent work.[595] Focusing the FM cases and particularly the prediction of associated $T_C$, it is found that only 53% of 786 compounds predicted to be FM,[500] are actually stable against AFM configurations generated automatically based on the method developed in Ref. 84. In this regard, the parameterization of the Heisenberg exchange parameters should be scrutinized, where explicit comparison of the total energies for various magnetic configurations can be valuable. Follow-up Monte Carlo simulations based on fitted exchange parameters reveal 26 materials out of 157 would exhibit $T_C$ higher than 400 K, and the results are modeled using ML with an accuracy of 73%. Unfortunately, the exact AFM ground states are not addressed, which is interesting for future investigation.

It is well known that 2D materials offer an intriguing playground for various topological phases particularly for magnetic materials including the QAHE[594], AFM TI[596], and semimetals.[417] HTP calculations based on the spin-orbit spillage[585] has been carried out to screen for magnetic and nonmagnetic topological materials in 2D materials,[597] resulting in four insulators with QAHE and seven magnetic semimetals. Such calculations are done on about 1000 compounds in the JARVIS database (https://www.ctcms.nist.gov/∼knc6/JVASP.html), and we suspect that more extensive calculations together with screening on the magnetic ground state can be interesting.

Beyond the calculations across various crystal structure prototypes, HTP calculations on materials of the same structural type substituted by different elements provide also promising candidates and insights on the 2D magnetism.[598] For instance, Chittari *et al.* performed calculations on 54 compounds of the $MAX_3$ type, where M = V, Cr, Mn, Fe, Co, Ni, A = Si, Ge, Sn, and X = S, Se, Te. It is observed that most of them are magnetic, hosting HMs, narrow gap semiconductors, *etc.* Importantly, strain is found to be effective to tailor the competition between AFM and FM ordering in such compounds.[599] Similar observations are observed in in-plane ordered MXene (i-MXene),[600] which can be derived from the in-plane ordered MAX compounds with nano-laminated structures.[601] Moreover, our calculations on the 2D materials of the $AB_2$ (A being TM and B = Cl, Br, I) indicate that $FeI_2$ is a special case adopting the 2H-type structure while all the neighboring cases have the 1T-type.[602] This might be related to an exotic electronic state as linear combination of the $t_{2g}$ orbitals in $FeI_2$ driven by the strong SOC of I atoms.[603] We found that the band structure of such $AB_2$ compounds is very sensitive to the value of the effective local Coulomb interaction U applied on the $d$-bands of TM atoms. This is by the way a common problem for calculations on 2D (magnetic) materials, where the screening of Coulomb interaction behave significantly from that in 3D bulk materials.[604]

## VI. FUTURE PERSPECTIVES

### A. Multi-scale modelling

As discussed in previous sections, HTP calculations based on DFT are capable of evaluating the intrinsic physical properties, which usually set the upper limits for practical performance. To make more realistic pre-dictions and to directly validate with the experiments, multi-scale modelling is required. The first issue is to tackle the thermodynamic properties which can be obtained by evaluating the Gibbs free energy considering the contributions from the electronic, lattice, and spin degrees of freedom. Moreover, the applications of magnetic materials at elevated temperature dictate the numerical calculations of the magnetic properties at finite temperature as well, *e.g.*, MAE at finite temperature.[116] The electronic structure will also be renormalized driven by the thermal fluctuations, *e.g.*, temperature driven topological phase transition in $Bi_2Se_3$.[605]

The major challenge for multi-scale modelling is how to make quantitative predictions for materials with structural features of large length scales beyond the unit cells, including both topological defects (*e.g.*, domain wall) and structural defects (*e.g.*, grain boundary). Taking permanent magnets as an example, as indicated in Fig. 1, both the $M_r$ and $H_a$ cannot reach the corresponding theoretical limit of $M_s$ and MAE. MAE (represented by the lowest-order uniaxial anisotropy K) sets an upper limit for the intrinsic coercivity $2K/(\mu_0 M_s)$ as for single-domain particles via coherent switching,[606] whereas the real coercivity is given by $H_a = \alpha 2K/(\mu_0 M_s) - \beta M_s$, where $\alpha$ and $\beta$ are the phenomenological parameters, driven by complex magnetization switching processes mostly determined by the microstructures. It is noted that $\alpha$ is smaller than one (about 0.1-0.3) due to the extrinsic mechanisms such as inhomogeneity, misaligned grains, etc., and $\beta$ accounts for the local demagnetization effects.[607] This leads to the so-called Brown's paradox.[608] That is, the magnetization switching cannot be considered as ideally a uniform rotation of the magnetic moments in a single domain, but is significantly influenced by the extrinsic processes such as nucleation and domain wall pinning,[609] which are associated with structural imperfections at nano-, micro-, and macroscopic scales.[607] Therefore, in addition to screening for novel candidates for permanent magnets, the majority of the current research are focusing on how to overcome the limitations imposed by the microstructures across several length scales.[610,611]

Furthermore, the functionalities of ferroic materials are mostly enhanced when approaching/crossing the phase transition boundary, particularly for magneto-structural transitions of the first-order nature. Upon such structural phase transitions, microstructures will be developed to accommodate the crystal structure change, *e.g.*, formation of twin domains due to the loss of point-group symmetry. Moreover, most FOPT occur without long-range atomic diffusion, *i.e.*, of the martensitic type, leading to intriguing and complex kinetics. As indicated above, what is characteristic to the athermal FOPTs is the hysteresis, leading to significant energy loss converted to waste heat during the cooling cycle for MCE materials.[33] While hysteresis typically arises as a consequence of nucleation, in caloric materials it occurs primarily due to domain-wall pinning, which is the net result of long-range elastic strain associated with phase transitions of interest. When the hysteresis is too large, the reversibility of MCE can be hindered. Therefore, there is a great impetus to decipher the microstructures developed during the magneto-structural transitions, particularly the nucleation and growth of coexisting martensitic and austenite phases. For instance, it is observed recently that materials satisfying the geometric compatibility condition tend to exhibit lower hysteresis and thus high reversibility, leading to further stronger cofactor conditions.[612]

Therefore, in order to engineer magnetic materials for practical applications, multi-scale modelling is indispensable. To this goal, accurate DFT calculations can be performed to obtain essential parameters such as the Heisenberg exchange, DMI, SOT, and exchange bias, which will be fed into the atomistic[108] and micromagnetic modelling.[22] It is noted that such scale-bridging modelling is required to develop fundamental understanding of spintronic devices as well. For instance, for the SOT memristors based on the heterostructures composing AFM and FM materials,[275] the key problems are to tackle the origin of the memristor-like switching behavior driven possibly by the interplay of FM domain wall propagation/pinning with the randomly distributed AFM crystalline grains, and to further engineer the interfacial coupling via combinatorial material combinations for optimal device performance.

Given the fact that the computational facilities have been significantly improved nowadays, HTP predictions can be straightforwardly made, sometimes too fast in comparison to the more time-consuming experimental validation. In this regard, HTP experiments are valuable. Goll *et al.* proposed to use reaction sintering as an accelerating approach to develop new magnetic materials,[613] which provides a promising solution to verify the theoretical predictions and to further fabricate the materials. Such a HTP approach has also been implemented to screen over the compositional space, in order to achieve optimal MSME performance.[614] Nevertheless, what is really important is a mutual responsive framework combining accurate theoretical calculations and efficient experimental validation.

### B. Machine learning

Another emergent field is materials informatics[615] based on advanced analytical machine learning techniques, which can be implemented as the fourth paradigm[616] to map out the process-(micro)structure-property relationship and thus to accelerate the development of materials including the magnetic ones. As the underlying machine learning is data hungry, it is critical to curate and manage databases. Although there is increasing availability of materials database such as Materials Project,[8] OQMD,[617] and NOMAD,[10] the sheer lack of data is currently a limiting factor. For instance, the existing databases mentioned above compile mostly the chemical compositions and crystal structures, whereas the physical properties particularly the experimentally measured results are missing. In this regard, a recent work trying to collect the experimental $T_C$ of FM materials is very interesting, which is achieved based on natural language processing on the collected literature.[618]

Following this line, machine learning has been successfully applied to model the $T_C$ of FM materials, which is still a challenge for explicit DFT calculations as discussed above. Dam *et al.* collected the $T_C$ for 108 binary RE-3d magnets and the database is regressed using the random forest algorithm.[619] 27 empirical features are taken as descriptors with detailed analysis on the feature relevance. Two recent work[620,621] started with the AtomWork database[622] and used more universal chemical and structural descriptors. It is observed that the Curie temperature is mostly driven by the chemical composition, where compounds with polymorphs are to be studied in detail with structural features. The other properties such as MAE[623] and magnetocaloric performance[624] can also be fitted, which makes it very interesting for the future.

HTP calculations generate a lot of data which can be further explored using machine learning. This has been performed on various kinds of materials and physical properties, as summarized in a recent review.[625] Such machine learning modelling has also been applied on magnetic materials. For instance, based on the formation energies calculated, machine learning has been used to model not only the thermodynamic stability of the full Heusler,[526,626] half Heusler,[627] and quaternary Heusler[628] compounds, but also the spin polarization of such systems.[629]

In this sense, machine learning offers a straightforward solution to model many intrinsic properties of magnetic materials, but is still constrained by the lack of databases. For instance, the $T_C$ and $T_N$ are collected for about 10,000 compounds in the AtomWork database,[622] with significant uncertainty for compounds with multiple experimental values thus should be used with caution. In addition, both machine learning modellings of the $T_C$[620,621] are done based on the random forest algorithm, whereas our test using the Gaussian kernel regression method leads to less satisfactory accuracy. This implies that the data are quite heterogeneous. Although our modelling can distinguish the FM and AFM ground state, it is still unclear about how to predict exact AFM configuration, which requires a database of magnetic structures. To the best of our knowledge, there is only one collection of AFM structures available in the MAGNDATA database, where there are 1100 compounds listed with the corresponding AFM ground states.[589] Therefore, there is a strong impetus to sort out the information in the literature and to compile a database of magnetic materials.

Importantly, it is suspected that the interplay of machine learning and multi-scale modelling will create even more significant impacts on identifying the process-structure-property relationships. For instance, phase field modelling aided with computational thermodynamics can be applied to simulate the microstructure evolution at the mesoscopic scale for both bulk and interfaces,[630] leading to reliable process-structure mapping. As demonstrated in a recent work,[631] the structure-property connection for permanent magnets can also be established based on massive micromagnetic simulations. Thus, HTP calculations performed in a quantitative way can be applied to generate a large database where the process-structure-property relationship can be further addressed via machine learning, as there is no unified formalism based on math or physics to define such linkages explicitly. Additionally, machine learning interatomic potentials with chemistry accuracy have been constructed and tested on many materials systems,[632] which are helpful to bridge the DFT and molecular dynamics simulations at larger length scales. We believe an extension of such a scheme to parameterize the Heisenberg Hamiltonian (Eq. (IV B)) with on-top spin-lattice dynamics simulations[121,223] will be valuable to obtain accurate evaluation of the thermodynamic properties for magnetic materials.

From the experimental perspective, adaptive design (*i.e.*, active learning) based on surrogate models has been successfully applied to guide the optimization of NiTi shape memory alloys[633] and $BaTiO_3$-based piezoelectric materials.[634] Such methods based on Bayesian optimization or Gaussian process work on small sample sizes but with a large space of features, resulting in guidance on experimental prioritization with greatly reduced number of experiments. This method can hopefully be applied to optimizing the performance of the 2-14-1 type magnets, as the phase diagram and underlying magnetic switching mechanism based on nucleation is well understood. Furthermore, the machine learning techniques are valuable to automatize advanced characterization. Small angle neutron scattering is a powerful tool to determine the microstructure of magnetic materials, where machine learning can be applied to accelerate the experiments.[635] Measurements on the spectral properties such as x-ray magnetic circular dichroism (XMCD) can also be analyzed on-the-fly in an automated way based on Gaussian process modelling.[636] In this regard, the hyperspectral images obtained in scanning transmission electronic microscopy can be exploited by mapping the local atomic positions and phase decomposition driven by thermodynamics. That is, not only the local structures but also the possible grain boundary phases assisted with diffraction and EELS spectra can be obtained, as recently achieved in x-ray spectro-microscopy.[637,638]

## VII. SUMMARY

In conclusion, despite the magnetism and magnetic materials have been investigated based on quantum mechanics for almost a hundred years (if properly) marked by the discovery of electron spin in 1928, it is fair to say that they are not under full control of us due to the lack of thorough understanding, as we are facing fundamental developments marked by progresses on AFM spintronics, magnetic topological materials, and 2D magnets, and we are exposed to complex multi-scale problems in magnetization reversal and magneto-structural phase transitions. Thus, we believe systematic calculations based on the state-of-the-art first-principles methods on an extensive list of compounds will at least get the pending issues better defined.

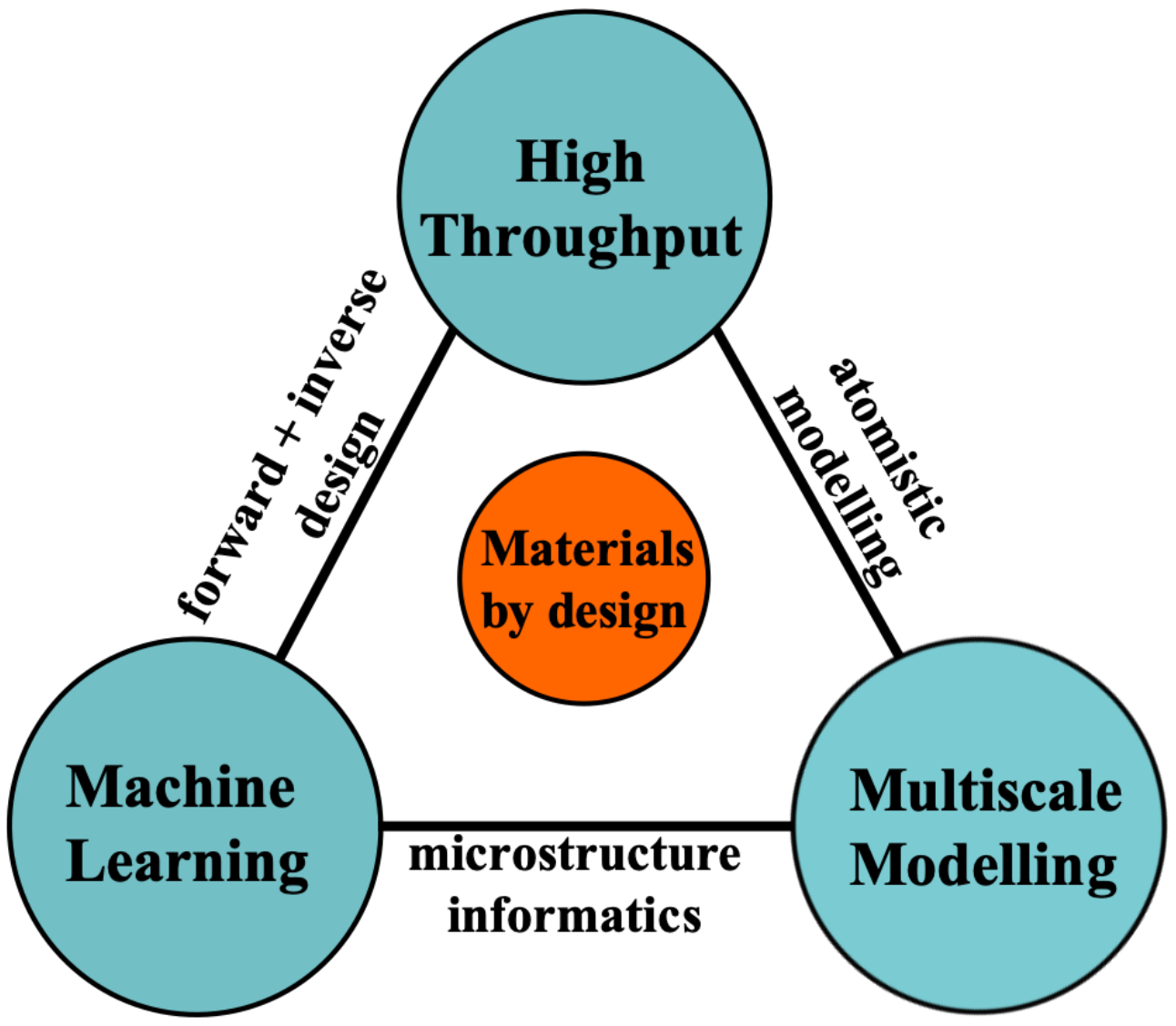

FIG. 11: Theoretical framework for future materials design

Among them, we would emphasize a few urgent and important aspects as follows,

- integrating the magnetic ground state searching into the available HTP platforms so that automated workflow can be defined. There are several solutions developed recently as discussed in Sect. IV C;
- implementation of a feasible DFT+DMFT framework for medium- (if not high-) throughput calculations. Two essential aspects are the interplay of Coulomb interaction, SOC, and hybridization and the treatment of magnetic fluctuations and paramagnetic states. This is also important for 2D magnets with enhanced quantum and thermal fluctuations;
- consistent framework to evaluate the transport properties which can be achieved based on accurate tight-binding-like models using Wannier functions;[639]
- quantitative multi-scale modelling of the magnetization reversal processes and the thermodynamic properties upon magneto-structural transitions in order to master hysteresis for permanent magnets and magnetocaloric materials;
- curation and management of a flexible database of magnetic materials, ready for machine learning.

Additionally, a common problem not only applicable for magnetic materials but also generally true for the other classes of materials subjected to property-optimization is a consistent way to consider substitution in both the dilute and concentrated limits. There are widely used methods such as virtual crystal approximation, coherent potential approximation,[640] special quasirandom structures,[117] cluster expansion,[641] *etc.*, but we have not yet seen HTP calculations done systematically using such methods to screen for substitutional optimization of magnetic materials.

Beyond the bare HTP design, we envision a strong interplay between HTP, machine learning, and multi-scale modelling, as sketched in Fig. 11. Machine learning is known to be data hungry, where HTP calculations based on DFT can provide enough data. In addition to the current forward predictions combining HTP and ML,[625] we foresee that inverse design can be realized,[642] as highlighted by a recent work on prediction new crystal structures.[643] As mentioned above, the marriage between HTP and multi-scale modeling helps to construct accurate atomistic models and thus to make the multi-scale modelling more quantitative and predictive. Last but not least, it is suspected that the interaction between machine learning and multi-scale modelling will be valuable to understand the microstructures, dubbed as microstructure informatics.[644] Such a theoretical framework is generic, *i.e.*, applicable to not only magnetic materials but also the other functional and structural materials.

All in all, HTP calculations based on DFT are valuable to provide effective screening on interesting achievable compounds with promising properties, as demonstrated for magnetic cases in this review. As the HTP methodology has been mostly developed in the last decade, we have still quite a few problems to be solved in order to get calculations done properly for magnetic materials, in order to be predictive. Concomitant with such an evolution of applying DFT to tackle materials properties, *i.e.*, with a transformation from understanding one compound to defining a workflow, more vibrant research on (magnetic) materials are expected, particularly combined with the emergent machine learning and quantitative multi-scale modelling.

**Acknowledgements** We appreciate careful proofreading and constructive suggestions from Manuel Richter, Ingo Opahle, and Nuno Fortunato. This work was supported by the Deutsche Forschungsgemeinschaft (DFG, German Research Foundation) Project-ID 405553726 TRR 270. Calculations for this research were conducted on the Lichtenberg high performance computer of TU Darmstadt.